\documentclass{rsproca_new}  

\usepackage{graphicx, cite}          
\usepackage{tikz}
\usepackage{amsmath,amssymb,amsfonts}
\usepackage{dsfont}
\usepackage{upgreek}
\usepackage{mathtools,mathdots}
\usepackage{xparse}
\usepackage{nameref}
\usepackage{graphicx}
\usepackage{caption}
\usepackage{subcaption}
\usepackage{float}
\usepackage{nicefrac}
\usepackage{multirow}
\usepackage{multicol}
\usepackage{hyperref}
\usepackage{multimedia}
\usepackage{arydshln}

\usepackage{algorithm}
\usepackage{algpseudocode}

\algrenewcommand\alglinenumber[1]{#1:}

\usepackage{pbox}

\newcommand{\defined}{\vcentcolon=}

\newcommand{\reals}{\mathds{R}}

\newcommand{\Mat}[1][]{\ifthenelse{\equal{#1}{}}{\text{Mat}}{\text{Mat}(#1)}}
\newcommand{\SO}[1]{\textnormal{SO}({#1})}
\newcommand{\SE}[1]{\textnormal{SE}({#1})}
\renewcommand{\so}[1]{\mathfrak{so}(#1)}
\newcommand{\se}[1]{\mathfrak{se}(#1)}

\newcommand{\TSE}[2][]{
	\ifthenelse{\equal{#1}{}}
	{{T\SE{#2}}}
	{{T_{#1}\SE{#2}}}
}
\newcommand{\dualTSE}[2][]{
	\ifthenelse{\equal{#1}{}}
	{{T^*\SE{#2}}}
	{{T^*_{#1}\SE{#2}}}
}

\newcommand{\dif}{\mathrm{d}}
\newcommand{\gradient}{\nabla}

\newcommand{\set}[1]{\left\{#1\right\}}

\newcommand{\abs}[1]{\left|#1\right|}
\newcommand{\norm}[1]{\left\lVert#1\right\rVert}

\newcommand{\inverse}{{-1}}

\newcommand{\transpose}{\top}

\newcommand{\Trace}[1]{\text{Tr}(#1)}

\newcommand{\inner}[2]{\left\langle #1, #2 \right\rangle}
\newcommand{\liebracket}[2]{\left[ #1, \ #2 \right]}
\newcommand{\Frobenius}{\text{F}}

\newcommand{\Vector}[1]{{#1}}
\newcommand{\Matrix}[1]{\mathrm{#1}}
\newcommand{\identity}[1]{\Matrix{I}_{#1}}
\newtheorem{proposition}{Proposition}
\newtheorem{definition}{Definition}
\newtheorem{remark}{Remark}
\def\R{{\mathds{R}}}
\def\0{{\mathbb{0}}}
\def\1{{\mathds{1}}}

\newcommand{\density}{\rho}
\newcommand{\radius}{r}
\newcommand{\area}{A}
\newcommand{\areamoment}{\Matrix{J}}
\newcommand{\position}{x}
\newcommand{\positions}{\Vector{\position}}
\newcommand{\director}{\mathsf{d}}
\newcommand{\orientation}{\Matrix{Q}}
\newcommand{\strains}{{\Vector{\varepsilon}}}
\newcommand{\strain}{\varepsilon}
\newcommand{\curvatures}{\Vector{\kappa}}
\newcommand{\curvature}{\kappa}
\newcommand{\shears}{\Vector{\nu}}
\newcommand{\shear}{\nu}
\newcommand{\internalforce}{n}
\newcommand{\internalforces}{\Vector{n}}

\newcommand{\internalcouples}{\Vector{m}}
\newcommand{\linearvelocity}{v}
\newcommand{\linearvelocities}{\Vector{v}}
\newcommand{\angularvelocity}{\omega}
\newcommand{\angularvelocities}{\Vector{\angularvelocity}}
\newcommand{\velocity}{w}
\newcommand{\velocities}{\Vector{\velocity}}
\newcommand{\momentum}{p}
\newcommand{\momentums}{\Vector{\momentum}}

\newcommand{\state}{\Matrix{q}}
\newcommand{\pose}{\Matrix{q}}
\newcommand{\strainmatrix}{\hat{\strains}}
\newcommand{\velocitymatrix}{\hat{\velocities}}

\newcommand{\costate}{\Matrix{\lambda}}
\newcommand{\transformedcostate}{\Matrix{K}}
\newcommand{\transformedcostateA}{\Matrix{K}_\text{A}}
\newcommand{\transformedcostateB}{\Matrix{K}_\text{B}}
\newcommand{\transformedcostateC}{\Matrix{K}_\text{C}}
\newcommand{\transformedcostateD}{\Matrix{K}_\text{D}}

\newcommand{\bendingrigidity}{\mathsf{B}}
\newcommand{\shearingrigidity}{\mathsf{S}}

\newcommand{\symmetricpart}{\Matrix{M}}

\newcommand{\Lagrangian}{\mathsf{L}}

\newcommand{\armlength}{L}

\newcommand{\hillsmodel}{h}
\newcommand{\contractileforce}{f}

\newcommand{\elastic}{\textnormal{e}}

\newcommand{\muscle}{\textnormal{m}}
\newcommand{\Gmuscle}{\mathsf{G}}
\newcommand{\desired}{\textnormal{d}}

\newcommand{\storedenergy}{W}
\newcommand{\potentialenergy}{\mathcal{V}}
\newcommand{\potential}{\mathcal{V}}
\newcommand{\kineticenergy}{\mathcal{T}}
\newcommand{\Hamiltonian}{\mathcal{H}}
\newcommand{\staticHamiltonian}{\mathsf{H}}

\newcommand{\inertia}{\mathsf{M}}
\newcommand{\intrinsic}{\circ}

\newcommand{\cycle}{\text{c}}

\newcommand{\damping}{\zeta}

\newcommand{\musclepositions}[1][]{\positions^{\ifthenelse{\equal{#1}{}}{\muscle}{#1}}}
\newcommand{\musclerelativepositions}[1][]{\Vector{\gamma}^{\ifthenelse{\equal{#1}{}}{\muscle}{#1}}}

\newcommand{\musclelength}[1][]{\ell^{\ifthenelse{\equal{#1}{}}{\muscle}{#1}}}
\newcommand{\musclestrain}[1][]{\strain^{\ifthenelse{\equal{#1}{}}{\muscle}{#1}}}
\newcommand{\muscleshears}[1][]{\shears^{\ifthenelse{\equal{#1}{}}{\muscle}{#1}}}
\newcommand{\muscletangent}[1][]{\Vector{\mathsf{t}}^{\ifthenelse{\equal{#1}{}}{\muscle}{#1}}}
\newcommand{\muscleforce}[1][]{\internalforce^{\ifthenelse{\equal{#1}{}}{\muscle}{#1}}}
\newcommand{\maxmusclestress}[1][]{\sigma^{\ifthenelse{\equal{#1}{}}{\muscle}{#1}}}
\newcommand{\maxmuscleforce}[1][]{\internalforce^{\ifthenelse{\equal{#1}{}}{\muscle}{#1}}_\textnormal{max}}
\newcommand{\muscleforces}[1][]{\internalforces^{\ifthenelse{\equal{#1}{}}{\muscle}{#1}}}
\newcommand{\musclecouples}[1][]{\internalcouples^{\ifthenelse{\equal{#1}{}}{\muscle}{#1}}}
\newcommand{\muscleactivation}[1][]{u^{\ifthenelse{\equal{#1}{}}{\muscle}{#1}}}
\newcommand{\muscleactivations}{\Vector{u}}
\newcommand{\staticmuscleactivation}[1][]{\alpha^{\ifthenelse{\equal{#1}{}}{\muscle}{#1}}}
\newcommand{\staticmuscleactivations}{\Vector{\alpha}}
\newcommand{\musclestoredenergy}[1][]{W^{\ifthenelse{\equal{#1}{}}{\muscle}{#1}}}

\newcommand{\Object}{\mathsf{C}}

\newcommand{\TM}{\textnormal{TM}}
\newcommand{\LM}[1][]{\textnormal{LM}{\ifthenelse{\equal{#1}{}}{}{_{#1}}}}
\newcommand{\OM}[1][]{\textnormal{OM}{\ifthenelse{\equal{#1}{}}{}{_{#1}}}}

\graphicspath{{figures/}}

\usepackage{titlesec}
\renewcommand\thesubsection{\arabic{section}.\arabic{subsection}} 
\renewcommand\thesubsubsection{\arabic{section}.\arabic{subsection}.\arabic{subsubsection}} 

\begin{document}

\title{Energy Shaping Control of a Muscular Octopus Arm Moving in Three Dimensions} 

\author{
Heng-Sheng Chang$^{1,2}$, Udit Halder$^{2}$, Chia-Hsien Shih$^{1}$, Noel Naughton$^{3}$, Mattia Gazzola$^{1,4,5}$, and Prashant G. Mehta$^{1,2}$}

\address{$^{1}$Department of Mechanical Science and Engineering, $^{2}$Coordinated Science Laboratory, 
$^{3}$Beckman Institute for Advanced Science and Technology, 
$^{4}$National Center for Supercomputing Applications, and
$^{5}$Carl R. Woese Institute for Genomic Biology, University of Illinois at Urbana-Champaign, IL, 61801, USA}

\subject{applied mathematics, biomechanics, differential equations, mathematical modeling}


\corres{Udit Halder\\
\email{udit@illinois.edu }}

\begin{abstract}                          
Flexible octopus arms exhibit an exceptional ability to coordinate large numbers of degrees of freedom and perform complex manipulation tasks.  As a consequence, these systems continue to attract the attention of biologists and roboticists alike.  In this paper, we develop a three-dimensional model of a soft octopus arm, equipped with biomechanically realistic muscle actuation.  Internal forces and couples exerted by all major muscle groups are considered.  An energy shaping control method is described to coordinate muscle activity so as to grasp and reach in 3D space. Key contributions of this paper are: (i) modeling of major muscle groups to elicit three-dimensional movements; (ii) a mathematical formulation for muscle activations based on a stored energy function; and (iii) a computationally efficient procedure to design task-specific equilibrium configurations, obtained by solving an optimization problem in the Special Euclidean group $\SE{3}$.  Muscle controls are then iteratively computed based on the co-state variable arising from the solution of the optimization problem. The approach is numerically demonstrated in the physically accurate software environment \textit{Elastica}. Results of numerical experiments mimicking observed octopus behaviors are reported.  
\end{abstract}
\maketitle

\section{Introduction} \label{sec:intro}

Interest in soft robots, specifically soft continuum arms (SCA), comes from their potential ability to perform complex tasks in unstructured environments as well as to operate safely around humans, with applications ranging from agriculture~\cite{uppalapati2020berry, chowdhary2019soft, laschi2012soft}
to surgery~\cite{cianchetti2018biomedical, wang2017cable, runciman2019soft}. An important bio-inspiration for SCAs is provided by octopus arms~\cite{li2012octopus, cianchetti2015bioinspired, grissom2006design, wu2021novel}. An octopus arm is hyper-flexible with nearly infinite degrees of freedom, seamlessly coordinated to generate a rich orchestra of motions such as reaching, grasping, fetching, crawling, or swimming~\cite{levy2017motor,kennedy2020octopus}.  How such a marvelous coordination is possible remains a source of mystery and amazement, and of inspiration to soft roboticists.  Part of the challenge comes from the intricate organization and biomechanics of the three major muscle groups---transverse, longitudinal, and oblique---which add to the overall complexity of the problem~\cite{kier1982functional, kier2007arrangement, feinstein2011functional, kier2016musculature}.


In this paper, we develop a bio-physical model of octopus arm equipped with virtual musculature, using the formalism of the Cosserat rod theory~\cite{antman1995nonlinear,gazzola2018forward}. In this type of modeling, a key concept is the \textit{stored energy function} of nonlinear elasticity theory whereby the internal forces and couples of a hyperelastic rod are obtained as the gradients of the stored energy function. The goal of this work is to extend the energy concept for following inter-related tasks: (i) Bio-physical modeling of the internal muscles, and (ii) Model-based control design.  The specific contributions on the two tasks are as follows.

\smallskip
\noindent 
\textbf{1. Muscle stored energy function.}  For each of the three major muscles of octopus arm, we present a mathematical model in terms of \textit{muscle stored energy function}.  Starting from first principle, and incorporating empirical force-length data~\cite{kier2002fast}, explicit formulae of the muscle stored energy function are derived.  Our approach represents the first such extension, to the continuum setting of a Cosserat rod, of the celebrated Hill's model for skeletal muscles \cite{hill1938heat, winter2009biomechanics, fung1996biomechanics}. This is contrasted with earlier work based on segment-based finite-dimensional models of octopus arm~\cite{yekutieli2005dynamic1, kang2012dynamic}.

\smallskip
\noindent
\textbf{2. Energy shaping control.} The energy-based modeling of muscles (actuators) greatly simplifies the ensuing control design.  It particular, it is shown to circumvent the key difficulty of computing the solution of the so called matching conditions during the application of energy shaping control~\cite{takegaki1981new,van2000l2,ortega2001putting}. 
In order to compute the muscle control input, the energy shaping control design is first posed as an optimization problem in the space of the special Euclidean group $\SE{3}$.  Its solution is obtained through an application of the maximum principle of optimal control theory, which also leads to a computationally tractable algorithm to compute the muscle control inputs.  Several novel features of the solution are noted, among which the physical interpretation of the co-state variable as resultant internal forces and couples exerted by activating the muscles.

\smallskip

While this paper focuses on the octopus arm, the above two contributions can broadly inform also the modeling and control of SCAs.  A key insight is that the conventional paradigm of modeling robot and actuators as two separate systems (in series) is no longer applicable for soft robots.  Worse yet, such an approach may be counter-productive because the resulting control problem is often computationally intractable.    
Because the actuators in SCAs---such as artificial muscles~\cite{mirvakili2018artificial, xiao2019biomimetic} or embedded pneumatic devices~\cite{singh2017constrained, singh2020designing}---are themselves mechanical systems, energy provides a unifying concept to simultaneously model {\em both} in an integrated manner. The total potential energy of an SCA is then simply the sum of its intrinsic elastic (passive) potential energy and the (active) potential energy of the actuator.  This renders the overall SCA a Hamiltonian control system for which energy shaping control methodology is a natural option with a rich history in robotics.  

The two contributions are numerically illustrated in the high-fidelity simulation package \textit{Elastica} \cite{gazzola2018forward,zhang2019modeling}. Results from two representative experiments, reaching and grasping,  are described.  These results help showcase the out-of-plane motion of the arm, mimicking observed octopus behaviors, e.g., an octopus twisting its arm to reach a target, and an octopus wrapping its arm around a cylindrical pole.  

Prior to this work, there have been a number of applications of the energy shaping control methodology to both rigid robots~\cite{holm2008kinetic, lin2019contact} and soft robots \cite{godage2014energy, gandhi2016energy, borja2022energy}. These demonstrations rely either on finite-dimensional articulated models (multiple rigid links) of the robot~\cite{franco2021energy, keppler2022underactuation} or a finite-dimensional approximation of the robot's energy~\cite{borja2022energy}.  In continuum settings, optimization-based formulations to design static equilibria appear in \cite{bretl2014quasi, till2017elastic}.  Here, we build on prior work in planar cases,   
whereby only longitudinal and transverse muscles were considered to obtain motions in 2D~\cite{chang2020energy,chang2021controlling, wang2021optimal, wang2022control, wang2022sensory}. While the planar modeling is an important first step, this paper represents a significant advancement including a comprehensive modeling and control of all major muscle groups to obtain complex 3D motions. 

The remainder of this paper is organized as follows.  An overview of the anatomy of the octopus arm musculature together with its kinematic modeling appears in \S\,\ref{sec:kinematic_model}. The dynamics of the arm, modeling of muscle actuation, and the control objectives appear in \S\,\ref{sec:dyamic_model}.  The details of the energy shaping control method are contained in \S\,\ref{sec:control}. Numerical experiments are reported in \S\,\ref{sec:numerics} and conclusions are given in \S\,\ref{sec:conclusion}.

\section{Kinematic model of an octopus arm and its musculature} 
\label{sec:kinematic_model}

\subsection{Pose of an arm}
A flexible octopus arm is modeled as a single Cosserat rod with rest length $\armlength$ in an inertial laboratory frame $\set{\mathsf{e}_1,\mathsf{e}_2,\mathsf{e}_3}$.  The arc-length coordinate of the center line of the arm is denoted as $s\in[0,\armlength]$.  Physically, the center line runs through the axial nerve cord of the octopus arm.   Along with $s$, the second independent coordinate is time $t$.  For notational ease, the partial derivatives with respect to $s$ and $t$ are denoted as $\partial_s:=\tfrac{\partial}{\partial s}$ and $\partial_t\defined\tfrac{\partial}{\partial t}$, respectively. 

The shape of the arm is described through its kinematic \textit{pose}  
\begin{equation*}
		\pose (s,t) :=\begin{bmatrix}
		\orientation (s,t) &\positions(s,t)\\
		0&1
		\end{bmatrix},\quad 0\leq s\leq \armlength, \quad t\geq 0
	\end{equation*}
where $\positions (s,t) \in \reals^3$ is the position vector of the center line at the material point $s$, and $\orientation(s,t) = \left[\director_1(s,t) ~\director_2(s,t) ~\director_3(s,t)\right] $ is a special orthogonal matrix whose column vectors are referred to as the director vectors.  The normal and binormal directors $\director_1$ and $\director_2$, respectively, span the cross section of the arm and $\director_3 = \director_1 \times \director_2$, where $\times$ denotes the vector cross product.  The matrix $\orientation(s,t)$ describes the orientation of the material frame at point $s$ at time $t$. See Fig.~\ref{fig:model}a for an illustration.

\subsection{Physiology and geometry of the muscles}

\begin{figure*}[!t]
	\centering
	\includegraphics[width=\textwidth, trim = {0 0 0 0}, clip = false]{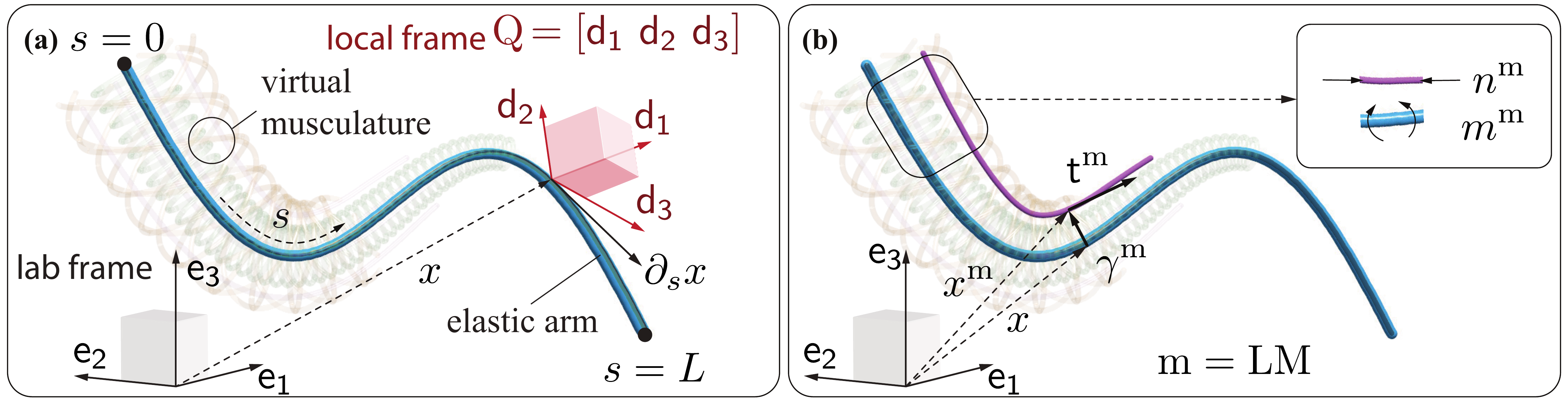}
	\caption{An octopus arm is modeled as a Cosserat rod. (a) The blue line denotes the center line of the arm where the base is denoted as $s=0$, and the tip is denoted as $s=\armlength$. At each time $t$, its position at arc-length location $s$ is denoted as $\positions(s,t)$ and its orientation (local frame) is $\orientation(s,t)$. (b) Example of a longitudinal muscle $\muscle=\LM$. The purple line denotes the position $\positions^\muscle$ of the virtual longitudinal muscle. The muscle exerts an active force $\muscleforces$ along the direction of the tangent vector $\muscletangent$. This results in a couple $\musclecouples$ acting on the center line.}
	\label{fig:model}
\end{figure*}

\begin{figure*}[!t]
	\centering
	\includegraphics[width=\textwidth]{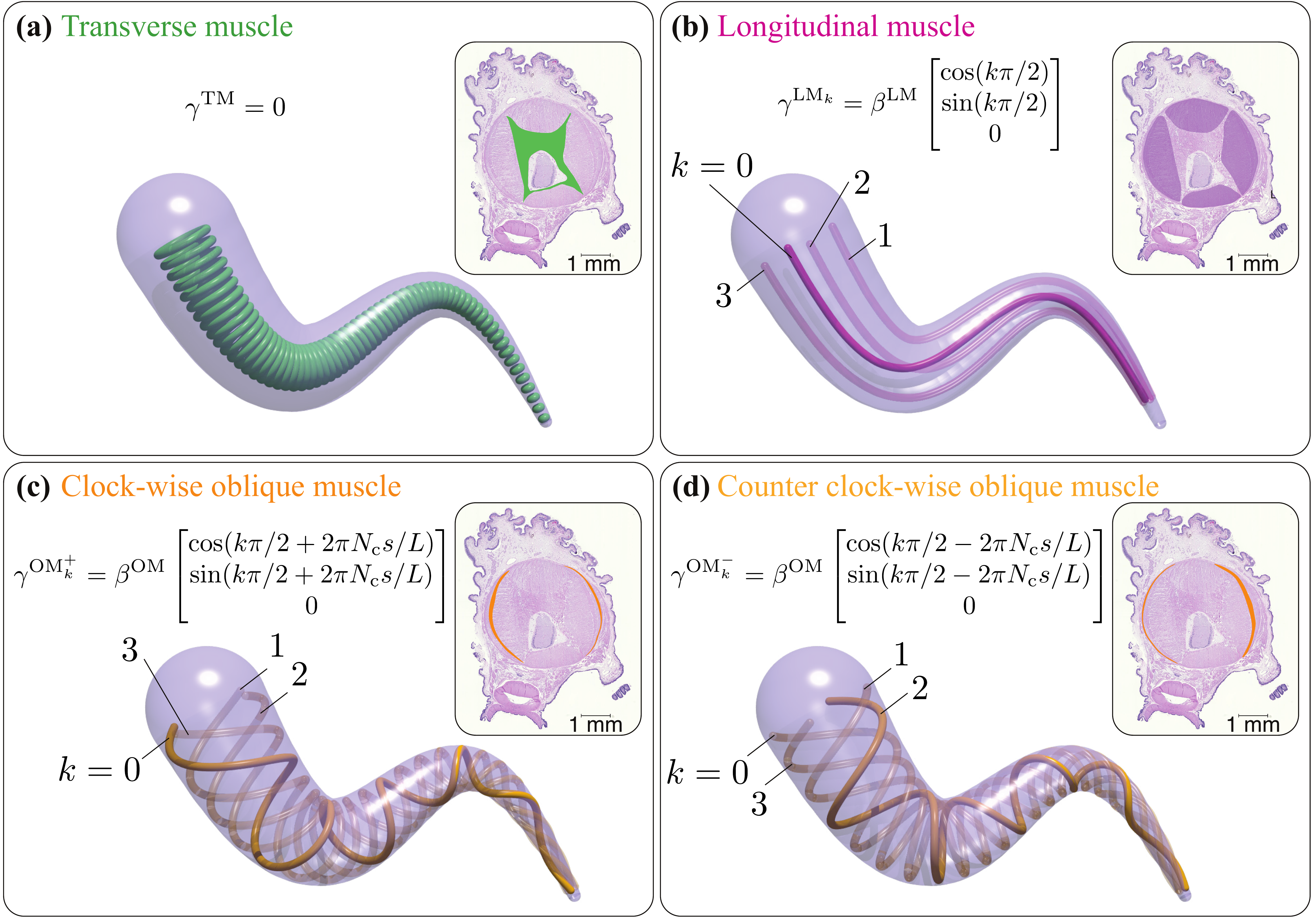}
	\caption{Illustration of the (virtual) musculature modeling.  Insets show a physiological cross section of an \textit{Octopus rubescens} arm (see~\ref{appdx:muscle_design} for details). Different muscle groups are color coded differently. Each subfigure shows the position of each type of muscle. Position vector of a muscle relative to the center line is denoted by $\musclerelativepositions$. (a) Transverse muscles are illustrated as green rings surrounding the axial nerve cord. Since they do not result in couples acting on the center line, their relative position is set as $\musclerelativepositions[\TM]=0$. (b) Longitudinal muscles are colored in purple. They run parallel to the center line. (c) and (d) are oblique muscles with clock-wise and counter clock-wise rotation, respectively. They are colored in orange. }
	\label{fig:muscle_model}
	\vspace*{-10pt}
\end{figure*}

In an octopus arm, the axial nerve cord is surrounded by densely packed muscles and connective tissue. Organizationally, there exist three major muscle groups~\cite{kier2007arrangement, feinstein2011functional, kier2016musculature} as seen from physiological cross-section of an \textit{Octopus rubescens} arm (Fig.~\ref{fig:muscle_model}):
\begin{enumerate}
\item Transverse muscles (TM) surround the nerve cord and are oriented orthogonally to the axial direction of the arm. 
\item Longitudinal muscles (LM) run parallel to the nerve cord.  
\item Oblique muscles (OM) are helically arranged around the longitudinal and transverse muscles.
\end{enumerate} 
While each of these muscles can only contract when activated, by virtue of their geometrical arrangement, internal forces (both contractile and extensional) and couples are generated, allowing  
the arm to shorten and bend (LM), extend (TM), and twist (OM).  
A generic muscle is denoted as $\muscle \in \set{\TM, \LM, \OM}$.  In the laboratory frame, its position is
		\begin{equation}
			\musclepositions:=\positions+\orientation\musclerelativepositions
			\label{eq:muscle_positions}
		\end{equation}
where $\musclerelativepositions$ is the muscle's relative position from the center line, represented in the material frame. The unit tangent vector of a muscle is defined as
\begin{align*}
\muscletangent\defined\orientation^\transpose\frac{\partial_s\musclepositions}{\abs{\partial_s\musclepositions}}
\end{align*}
An illustration of these quantities is given in Fig.~\ref{fig:model}b for $\muscle = \LM$. For each muscle group, corresponding locations are sketched in Fig.~\ref{fig:muscle_model}, tabulated in Table~\ref{tab:muscle_model}, and further detailed in \ref{appdx:muscle_design}.

\begin{table}[!t]
	\footnotesize
	\centering
	\caption{Nomenclature}
	\hspace*{-5pt}
	\begin{tabular}{ll | ll | ll}
		\rowcolor{black}
		\multicolumn{4}{c}{\color{white} Arm related variables} & 
		\multicolumn{2}{c}{\color{white} Generic muscle related variables}\\
		$\positions$ & center line position & $\armlength$ & rest length of the arm & $\musclepositions$ & muscle position\\
		$\director_i$ & center line directors & $\velocities$ &  velocity & $\musclerelativepositions$ & muscle relative position\\
		$\orientation$ & center line orientation & $\linearvelocities$ &  linear velocity & $\muscletangent$ & muscle tangent director \\
		$\pose$ &  pose & $\angularvelocities$ &  angular velocity & $\muscleshears$ & muscle strain\\
		$\strains$ &  strain & $\momentums$ &  momentum & $\musclelength$ & muscle local length\\
		$\shears$ &  shears and stretch & $\internalforces$ &  internal forces  & $\muscleforces$ & muscle forces\\
		$\curvatures$ & curvatures and twist & $\internalcouples$ & internal couples & $\musclecouples$ & muscle couples\\
		$\radius$ & radius of the cross section & $\rho$ &density of the arm & $\muscleactivation$ & muscle activation\\
		\hline
	\end{tabular}
	\label{tab:nomenclature}
\end{table}

\colorlet{shadecolor}{gray!40}
\begin{table*}[!tb]
	\caption{Summarized muscle model}
	\hskip-10pt
	\centering
	\begin{tabular}{c ;{3.14159pt/2.71828pt} cc ;{3.14159pt/2.71828pt} ccc} 
		\rowcolor{black}
		\hline
		& \multicolumn{2}{c}{\color{white} Muscle model notation} &  \multicolumn{3}{c}{\color{white} Values of parameters}\\
		\rowcolor{shadecolor}
		&&&&&\\[-0.8em]
		\rowcolor{shadecolor}
		muscle & relative position & local length &  & max stress & normalized area  \\
		\rowcolor{shadecolor}
		$\muscle$ & $\musclerelativepositions$ & $\musclelength$ & \multirow{-2}{*}{$\dfrac{\beta^\muscle}{\radius}$} & $\maxmusclestress_{\text{max}}$  [kPa] & $\area^\muscle/\area$ \\
		&&&&&\\[-0.6em]
		$\TM$ & $0$ & $\sqrt{\dfrac{\abs{\muscleshears[\TM]_\intrinsic}}{\abs{\muscleshears[\TM]}}}$ & - & $15$ & $\dfrac{1}{8}$ \\
		&&&&&\\[-0.6em]
		&&&&&\\[-0.6em]
		$\LM[k]$ & $\beta^{\LM}\begin{bmatrix}\cos(k\pi/2)\\\sin(k\pi/2)\\0\end{bmatrix}$ & $\dfrac{\abs{\muscleshears[{\LM[k]}]}}{\abs{\muscleshears[{\LM[k]}]_\intrinsic}}$ & $\dfrac{5}{8}$ & $10$ & $\dfrac{1}{16}$ \\
		&&&&&\\[-0.6em]
		&&&&&\\[-0.6em]
		$\OM^+_k$ & $\beta^{\OM}\begin{bmatrix}\cos(k\pi/2+2\pi N_\cycle s/\armlength)\\\sin(k\pi/2+2\pi N_\cycle s/\armlength)\\0\end{bmatrix}$ & $\dfrac{\abs{\muscleshears[{\OM^+_k}]}}{\abs{\muscleshears[{\OM^+_k}]_\intrinsic}}$ & $\dfrac{15}{16}$ & $100$ & $\dfrac{1}{256}$ \\
		&&&&&\\[-0.6em]
		&&&&&\\[-0.6em]
		$\OM^-_k$ & $\beta^{\OM}\begin{bmatrix}\cos(k\pi/2-2\pi N_\cycle s/\armlength)\\\sin(k\pi/2-2\pi N_\cycle s/\armlength)\\0\end{bmatrix}$& $\dfrac{\abs{\muscleshears[{\OM^-_k}]}}{\abs{\muscleshears[{\OM^-_k}]_\intrinsic}}$ & $\dfrac{15}{16}$ & $100$ & $\dfrac{1}{256}$ \\
		&&&&&\\[-0.6em]
		\hline
	\end{tabular}
	\label{tab:muscle_model}
\end{table*}		

\subsection{Strains, velocities, and energies}

For each fixed $(s,t)$, the pose $\pose(s,t)$ is an element of the special Euclidean group $\SE{3}$.  The orientation $\orientation(s,t)$ is an element of the special orthogonal group $\SO{3}$.  The associated Lie algebra are denoted as $\se{3}$ and $\so{3}$, respectively.  
A linear map $[\cdot]_\times ~:~ \reals^3 \rightarrow \so{3}$ is defined as
\begin{align*}
[\Vector{z}]_\times \defined \begin{bmatrix*} 0 & -z_3 & z_2 \\ z_3 & 0 & -z_1 \\ -z_2 & z_1 & 0 \end{bmatrix*},\quad\Vector{z}\in\reals^3
\end{align*}
and its inverse map is denoted as $\text{vec}[\cdot] ~:~ \so{3} \rightarrow \reals^3$.  
The strains and the velocities of the arm are defined as follows 
\begin{equation} 
		\partial_s\pose(s,t)=\pose(s,t)\strainmatrix(s,t),\quad 
		\strainmatrix \defined \begin{bmatrix*}[c] [\curvatures]_{\times} & \shears \\ 0 & 0 \end{bmatrix*}
		\label{eq:kinematics}
\end{equation}
where $\strainmatrix(s,t)\in\se{3}$ with $\strains\defined ( \shears,\curvatures) \in \reals^6$ referred to as \textit{strain}. The strain includes shears $(\shear_1,\shear_2)$, stretch $\shear_3$, curvatures $(\curvature_1, \curvature_2)$, and twist $\curvature_3$.  Similarly,   
	\begin{equation}\label{eq:equation_of_motions}
		\partial_t \pose(s,t)=\pose(s,t)\velocitymatrix(s,t),\quad 
		\velocitymatrix\defined\begin{bmatrix*}[c] [\angularvelocities]_{\times} & \linearvelocities \\ 0 & 0 \end{bmatrix*}
	\end{equation}
	where $\velocitymatrix(s,t)\in\se{3}$ with $\velocities=(\linearvelocities,\angularvelocities) \in \reals^6$ referred to as \textit{velocity}. Note that elements of $\strains$ and $\velocities$ are strains and velocities expressed in the material frame.

The \textit{kinetic energy} of the arm is given by
 \begin{equation}\label{eq:kinetic_energy}
 \kineticenergy (\momentums) = \int_0^{\armlength} 
 																\frac{1}{2} \abs{p}^2_{\inertia^{-1}}~\dif s
\end{equation}
where $\inertia$ is the inertia matrix (see \ref{appdx:parameters}) and $\momentums\defined\inertia\velocities$ is the \textit{momentum} in the material frame. The vector norm is defined as $\abs{\Vector{z}}_\Matrix{A}^2\defined\Vector{z}^\transpose\Matrix{A}\Vector{z}$, for any vector $\Vector{z}\in\reals^n$, and $\Matrix{A}$ is a symmetric positive definite $n\times n$ real matrix. The \textit{elastic potential energy} is defined as
\begin{align}
	\potential^\elastic (\pose) &= \int_0^{\armlength} W^\elastic \left(\strains \right) \dif s 
\end{align}
where the function $W^\elastic : \reals^6 \rightarrow \reals$ is the elastic \textit{stored energy function} of the rod. Assuming a \textit{hyperelastic} rod, the elastic internal forces and couples $( \internalforces^\elastic,\internalcouples^\elastic)$ are gradients of this function, leading to the \textit{constitutive relationships}
\begin{align}
\internalforces^\elastic  = \frac{\partial W^\elastic}{\partial \shears} (\strains),\quad\internalcouples^\elastic = \frac{\partial W^\elastic}{\partial \curvatures} (\strains)
\label{eq:passive_elasticity}
\end{align}

\begin{remark}
The simplest example of the elastic stored energy function is the quadratic form
\begin{align}\label{eq:elastic_potential_energy}
W^\elastic(\strains) = \frac{1}{2} \left( \abs{\shears - \shears^\intrinsic}_\shearingrigidity^2+\abs{\curvatures - \curvatures^\intrinsic}_\bendingrigidity^2 \right)
\end{align}
where $\shearingrigidity,\bendingrigidity$ are the shearing and bending rigidity matrices of the rod (see \ref{appdx:parameters}) , respectively, and $\strains^\intrinsic = ( \shears^\intrinsic,\curvatures^\intrinsic)$ is the intrinsic strain of the rod. This particular form of the elastic stored energy function yields the linear stress-strain constitutive relation
\begin{align*}
\internalforces^\elastic = \frac{\partial W^\elastic}{\partial \shears} = \shearingrigidity (\shears - \shears^\intrinsic),\quad\internalcouples^\elastic = \frac{\partial W^\elastic}{\partial \curvatures} = \bendingrigidity (\curvatures - \curvatures^\intrinsic)
\end{align*}
\end{remark}

\section{Mathematical model of the control problem}
\label{sec:dyamic_model}

\subsection{Dynamics}
The dynamics of a Cosserat rod are captured by the system of nonlinear partial differential equations \cite{antman1995nonlinear, gazzola2018forward}
\begin{subequations}\label{eq:dynamics}
		\begin{align}
			\partial_t \pose &= \pose \velocitymatrix \\
			\partial_t \momentums
				&= \begin{bmatrix} \partial_s \internalforces+\curvatures\times\internalforces -\angularvelocities\times(\density\area\linearvelocities)\hfill \\
					\partial_s \internalcouples + \curvatures\times\internalcouples +  \shears \times \internalforces-\angularvelocities\times(\density\areamoment\angularvelocities) \end{bmatrix} + \begin{bmatrix} \mathsf{f} \\ \mathsf{c} \end{bmatrix}	  - \damping\momentums					
		\end{align}
	\end{subequations}
where $(\internalforces,\internalcouples)$ are internal forces and couples, represented in the material frame,  $(\mathsf{f}, \mathsf{c})$ are external body forces and couples in the material frame, and $\damping$ is the damping matrix which serves as a reduced-order model (see \ref{appdx:parameters}) for the viscoelastic properties of the elastic arm \cite{gazzola2018forward, chang2020energy}. The rod dynamics \eqref{eq:dynamics} (in the absence of external body forces and couples) are accompanied by a set of fixed-free type boundary conditions 
\begin{equation}
		\positions (0,t)= \positions_0,~\orientation(0,t)=\orientation_0,~\internalforces(\armlength,t)=0,~\internalcouples(\armlength,t)=0, ~~ \forall t\geq 0
		\label{eq:boundary_conditions}
\end{equation}
where $\positions_0 \in \R^3$ and $\orientation_0 \in \SO{3}$ are given position and orientation at the base, respectively. 
The control (muscles) affects the dynamics by modifying the overall internal forces and couples $( \internalforces,\internalcouples)$ as 
\begin{align}
\internalforces = \internalforces^\elastic + \sum_\muscle \muscleforces,\quad\internalcouples = \internalcouples^\elastic + \sum_\muscle \musclecouples 
\label{eq:internal_forces}
\end{align}
where $(\internalforces^\elastic, \internalcouples^\elastic)$ are restoring loads due to passive elasticity (see \eqref{eq:passive_elasticity}) and $(\muscleforces , \musclecouples)$ are active loads on the arm resulting from muscle contractions.  The latter depend on the muscle activations  $\muscleactivations := \{ \muscleactivation \}$ that serve as the control input.  
It remains to model the effect of muscles, namely $(\muscleforces , \musclecouples)$ as functions of the control $\muscleactivation$. 

\subsection{Forces and couples from muscle actuation} \label{sec:muscle_force_couple}

We model the arm and its musculature via a set of virtual muscles whose effects, upon contraction, are translated into resulting forces and couples acting on a single Cosserat rod, representing the arm itself. We start by noting that each muscle can only produce local contractile forces. 
According to the classical Hill's model \cite{hill1938heat, audu1985influence, fung1996biomechanics, winter2009biomechanics}, muscle forces are related to muscle lengths (relative to rest state)
$\musclelength$ through a force-length relationship described by a (normalized) function $\hillsmodel(\musclelength)\in [0,1]$. Descriptions of such force-length relationships are found in \cite{yekutieli2005dynamic1, chang2021controlling, zullo2022octopus}, and the specific form of $\hillsmodel(\cdot)$, fitted from experimental data \cite{kier2002fast}, is given in \S~\ref{sec:simulation_setup}. For simplicity, here we use the same force-length relationship for all muscle groups. However, different force-length relationships for different muscle groups \cite{zullo2022octopus} can be accommodated. Moreover, the muscle contractile force is proportional to its (rest) area $\area^\muscle (s)$, through the maximum force per unit area ($\sigma^\muscle_{\text{max}}$) that the muscle can provide. Hence, we write the active local muscle contractile force as
\begin{align}
\contractileforce^\muscle (s, \musclelength) = \sigma^\muscle_{\text{max}} \area^\muscle (s)  \hillsmodel(\musclelength)
\end{align}

In order to extend the Hill's model to express the effect of muscles on the single Cosserat rod representing the arm, we need to account for their geometric arrangement (Fig.~\ref{fig:muscle_model}).
Depending on the muscle group, active contractions result in contractile ($\LM, \OM$) or extensile ($\TM$) forces as well as couples ($\LM, \OM$) on the arm. For example, transverse muscle contractions create radial shortening which in turn makes the arm extend in length due to incompressibility. Thus, local $\TM$ contractile forces effectively generate an axial extensile force on the arm. The model describing these effects is provided next. 

For a generic muscle $\muscle$, the control input is the intensity of the actuation denoted as $\muscleactivation(s,t) \in [0,1]$ where $\muscle\in\set{\TM,\LM,\OM}$. 
Physically, the control input $\muscleactivation$ represents the innervating motor neuron stimulation (e.g. the firing frequency). The bound on the control is the manifestation of neurophysiological limits. When activated, the muscle exerts a resultant internal force $\muscleforces$ along the tangent direction of the muscle $\muscletangent$, modulated by $\muscleactivation$. 
That is 
		\begin{equation}
			\muscleforces\defined\pm\muscleactivation \contractileforce^\muscle \muscletangent
			\label{eq:muscle_forces}
		\end{equation}
where the negative sign indicates extensile force in the case of transverse muscles. Because the longitudinal and oblique muscles are located away from the center line, they also generate the couple
		\begin{align}
		\musclecouples\defined\musclerelativepositions \times \muscleforces
		\label{eq:muscle_couples}
		\end{align}
Notice that the transverse muscles are perpendicular to the arm axis and thus they do not produce resulting couples. Hence we set $\musclerelativepositions[\TM]= 0$ (see discussions in \ref{appdx:muscle_design}).
 		
It remains to specify the model for local muscle length $\musclelength$. We first define the muscle strain $\muscleshears$ by the following relationship
\begin{equation*}
		\partial_s \positions^{\muscle}=:\orientation\muscleshears
\end{equation*}
Recalling the definition of the muscle position vector $\musclepositions$ in \eqref{eq:muscle_positions}, we then readily obtain
		\begin{align}\label{eq:muscleshears}
			\muscleshears=\shears+\curvatures\times\musclerelativepositions+\partial_s\musclerelativepositions
		\end{align}
As a result, a natural definition of the intrinsic or rest muscle strain $\muscleshears_\intrinsic$ as a function of the intrinsic rod strain $(\shears^\intrinsic, \kappa^\intrinsic)$ is given by		
\begin{equation}
	\muscleshears_\intrinsic :=\shears^\circ+\curvatures^\circ\times\musclerelativepositions+\partial_s\musclerelativepositions
\end{equation}		
The local muscle length is then taken to be a function of the muscle strain, i.e. $\musclelength=\musclelength(|\muscleshears|)$. For all three types of muscles, the model for $\musclelength$ appears in Table~\ref{tab:muscle_model}. Explanations for such modeling are provided in \ref{appdx:muscle_design}.

\subsection{Control problem} \label{sec:control_problem}
It is useful to view the dynamics \eqref{eq:dynamics} as a Hamiltonian control system. The Hamiltonian of a non-actuated arm is the sum of kinetic and elastic potential energy.  Dynamics of an actuated arm are abstractly written as 
\begin{equation}\label{eq:control_system}
\frac{\dif}{\dif t}\begin{bmatrix}\pose\\\momentums\end{bmatrix}=\begin{bmatrix}
0&\Upxi \\-\Upxi^*&- \Upomega - \damping \inertia \end{bmatrix}\begin{bmatrix}\gradient_{\pose} \potentialenergy^{\elastic}\\\gradient_{\momentums} \kineticenergy\end{bmatrix} + \begin{bmatrix}0\\\mathsf{G}\Vector{u}\end{bmatrix}
\end{equation}
where the map $\Upxi$, its dual map $\Upxi^*$, the map $\Upomega$, and the generalized gradients $\gradient_{\pose}, \gradient_{\momentums}$ are used to express the dynamics \eqref{eq:dynamics} in the canonical form \cite{rashad2019port, van2013port}. 
The term $\Gmuscle \Vector{u}$ represents the overall forces and couples generated by the muscles. 
Explicit forms of  $\Upomega$ and $\Gmuscle \Vector{u}$ are given in \ref{appdx:parameters}.

The control problem is to obtain stabilizing control inputs $\Vector{u}$ for the system \eqref{eq:control_system} so as to achieve some predefined tasks. Here, the tasks are motivated by stereotypical motions observed in octopuses \cite{levy2017motor, kennedy2020octopus}. Two of these motions are reaching and grasping.  
Mathematically, these two problems are modeled by specifying an objective function for an optimization problem.  Details appear in \S\,\ref{sec:exp_reaching} and \S\,\ref{sec:exp_grasping}. 

\section{Energy shaping control method} \label{sec:control}

\subsection{Difficulties with the matching conditions}

Consider the Hamiltonian control system \eqref{eq:control_system} with its total energy or the Hamiltonian given by $\kineticenergy+\potentialenergy^{\elastic}$. The key idea of the (potential) energy shaping control scheme is to implement controls $\Vector{u}$ such that the Hamiltonian is modified from $\kineticenergy+\potentialenergy^{\elastic}$ to $\kineticenergy+\potentialenergy^\desired$, where $\potentialenergy^\desired$ is a desired potential energy.
That is, the controlled system evolves according to 
\begin{equation}\label{eq:hamilton_equations_desired}
\frac{\dif}{\dif t}\begin{bmatrix}\pose\\\momentums\end{bmatrix}=\begin{bmatrix}
0& \Upxi\\-\Upxi^*&- \Upomega - \damping \inertia\end{bmatrix}\begin{bmatrix}\gradient_{\pose} \potentialenergy^\desired\\\gradient_{\momentums} \kineticenergy\end{bmatrix}
\end{equation}
For the systems \eqref{eq:control_system} and \eqref{eq:hamilton_equations_desired} to \textit{match} (for arbitrary choice of initial conditions), it must follow that 
\begin{align}
\mathsf{G}\Vector{u} = \Upxi^*\left(\gradient_\pose\potentialenergy^{\elastic}-\gradient_\pose\potentialenergy^\desired\right), \quad \forall\; (\pose,\momentums)
\label{eq:equilibrium_match}
\end{align}
The operator $\Gmuscle$ is viewed as the actuator (muscle) constraint that restricts the muscle forces and couples to be applied only in certain directions (see~\ref{appdx:parameters}). In other words, the muscle group $\TM$, for example, can only produce axial forces on the arm. 
This is in contrast with the fully actuated case where $\Gmuscle$ is the identity operator, i.e. forces and couples on the arm are directly specified via arbitrary continuous functions and may be applied along any desired direction, as in \cite{chang2020energy}.
As a result, equation \eqref{eq:equilibrium_match} restricts the set of attainable desired potential energies $\potentialenergy^\desired$. 
The restriction on $\potentialenergy^\desired$ is expressed in the form of the matching condition which is described next.   

Let $\mathsf{G}^\perp$ denote the left annihilator of $\mathsf{G}$, i.e., $\inner{\mathsf{G}^{\perp}}{\mathsf{G}\Vector{u}}_{\mathrm{L}^2} = 0, \text{ for all}\ (\pose, \momentums)$ and for all $u$. Here $\inner{\cdot}{\cdot}_{\mathrm{L}^2}$ represents the usual $\mathrm{L}^2$ inner product. Then applying $\mathsf{G}^{\perp}$ to both sides of~\eqref{eq:equilibrium_match} yield
\begin{align}
\inner{\mathsf{G}^{\perp} }{\Upxi^*\left(\gradient_\pose\potentialenergy^{\elastic}-\gradient_\pose\potentialenergy^\desired\right)}_{\mathrm{L}^2} = 0, ~~ \forall \; (\pose, \momentums)
\label{eq:potential_matching}
\end{align}
Equation \eqref{eq:potential_matching} is referred to as the (potential) \textit{matching condition} in literature~\cite{bloch2000controlled, blankenstein2002matching,ortega2002stabilization,ortega2002interconnection}.  Its solution determines the set of allowable $\potentialenergy^\desired$ that may be used in the energy shaping control design.  For example, in the fully actuated case $\mathsf{G}^{\perp}=0$ so the matching condition is trivial and any choice of $\potentialenergy^\desired$ is a solution. However, in the presence of actuator constraints, the matching condition is impractical to solve, requiring convoluted and specialized techniques, e.g. the $\lambda$-method \cite{auckly2000control}, or additional assumptions \cite{bloch2000controlled, gomez2001stabilization}, even in finite dimensional settings. 


Here, we do not attempt to directly solve the matching condition. 
Instead, we show in \S~\ref{sec:muscle_energy_function} that it is possible to 
express active muscle forces and couples as gradients of a \textit{muscle stored energy function}. This implicitly solves the matching condition, leading to the energy shaping control law of \S~\ref{sec:energy_shaping_control_law}. 


\subsection{Muscle stored energy function}	\label{sec:muscle_energy_function}
Inspired by hyperelastic rod theory, we express the internal muscle forces and couples (see \eqref{eq:muscle_forces} and \eqref{eq:muscle_couples}) as gradients of a scalar \textit{muscle stored energy function}, defined as follows.
\begin{definition}
For a generic muscle $\muscle$, let the rest strain vector (shear and stretch) be denoted by $\Vector{\shear}^\muscle_\intrinsic$. Then, the muscle \textit{stored energy function} is defined as the line integral along a (piecewise) smooth curve $C \in \R^3$
\begin{equation}
			\musclestoredenergy(s,\shears^\muscle)\defined \left\{
\begin{aligned}
&\int_C~~\contractileforce^\muscle(s, \musclelength(\Vector{\abs{y}}))\frac{\Vector{y}}{\abs{\Vector{y}}}\cdot\dif\Vector{y}, &&\text{if } ~\muscle \in \{ \LM, \OM\} \\
&\int_C -\contractileforce^\muscle(s, \musclelength(\Vector{\abs{y}}))\frac{\Vector{y}}{\abs{\Vector{y}}}\cdot\dif\Vector{y}, &&\text{if } ~ \muscle = \TM
\end{aligned}\right.\label{eq:muscle_stored_energy}
\end{equation}
where $\Vector{y} : [0, 1] \mapsto C$ is a parameterization of the curve $C$, and  
$\Vector{y}(0)=\Vector{\shears}^\muscle_\intrinsic,~ \Vector{y}(1)=\Vector{\shears}^\muscle$. 
\end{definition}
Note that the integral of \eqref{eq:muscle_stored_energy} is path-independent as shown in~\ref{appdx:muscle_energy_proof}. Given the stored energy function \eqref{eq:muscle_stored_energy}, muscle forces and couples are obtained as follows.

\begin{figure*}[!t]
	\centering
	\includegraphics[width=\textwidth, trim = {0 300 0 300}, clip = false]{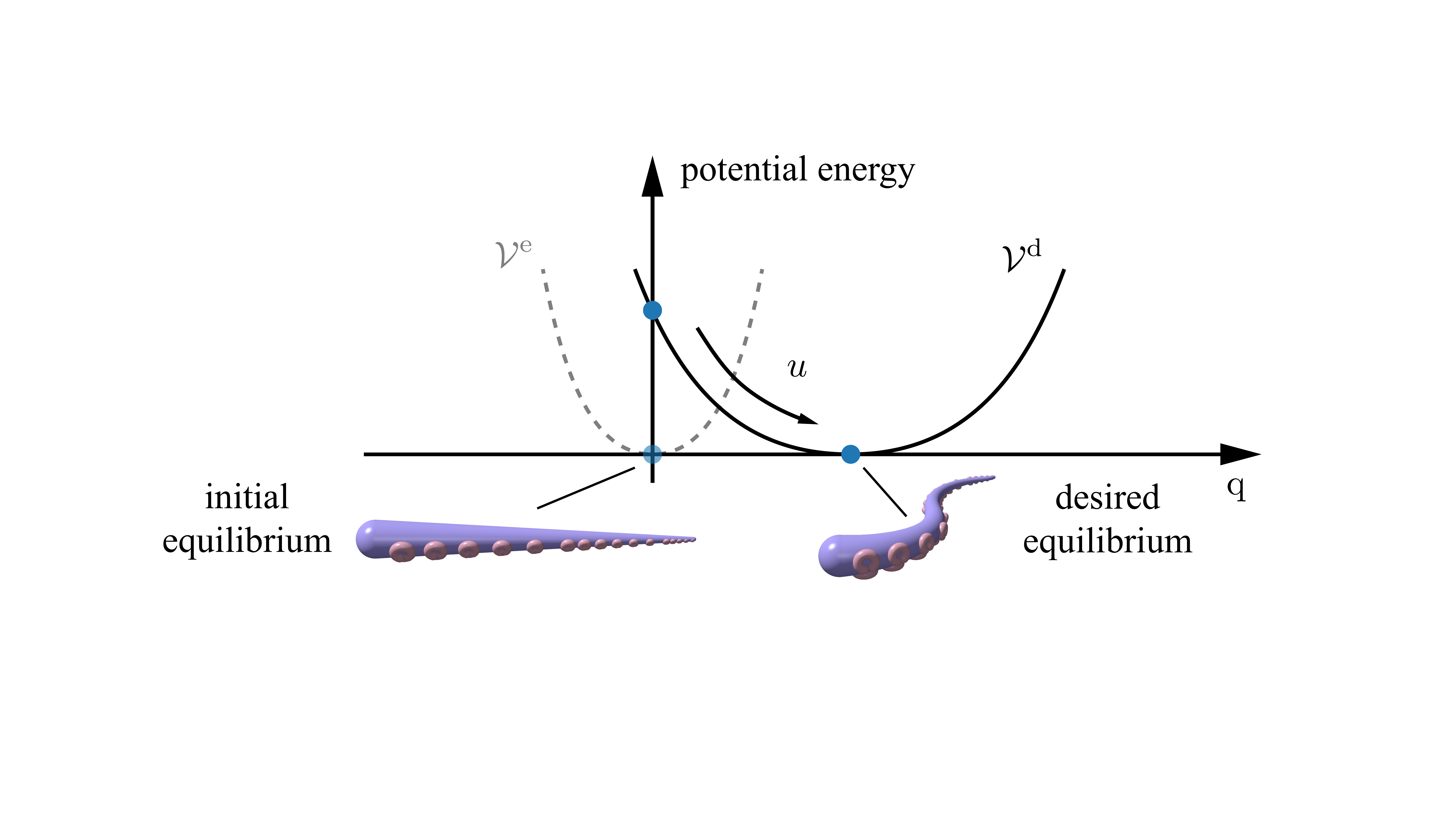}
	\caption{An illustration of the energy shaping control method. The system starts as a straight rod (equilibrium of potential energy landscape $\potentialenergy^\elastic$) and moves to a new equilibrium configuration based on the desired potential energy landscape $\potentialenergy^\desired$, through the control $\Vector{u}$.}
	\label{fig:energy_shaping}
\vspace{-10pt}
\end{figure*}

\begin{proposition} \label{prop:muscle_energy}
Suppose $\muscleactivation = \muscleactivation (s, t) $ is a given activation of a generic muscle $\muscle$. Then, resulting muscle forces and couples acting on the arm are given by
\begin{align}
	\muscleforces =\muscleactivation \frac{\partial \musclestoredenergy}{\partial\shears}, \quad \musclecouples =\muscleactivation \frac{\partial\musclestoredenergy}{\partial\curvatures} 
\label{eq:muscle_energy}
\end{align}
\end{proposition}

\begin{proof}
See ~\ref{appdx:muscle_energy_proof}.
\end{proof}

Proposition~\ref{prop:muscle_energy} leads to the following expression for the total stored energy function of the arm
\begin{align}
W(s, \strains; \Vector{u}) = W^\elastic (\strains) + \sum_\muscle u^\muscle W^\muscle (s, \muscleshears(\strains))
\label{eq:total_stored_energy}
\end{align}
Thus, the total potential energy of the arm becomes
\begin{align}
\potential (\pose; \Vector{u}) = \int_0^{\armlength} W(s, \strains(s); \Vector{u})~ \dif s
\label{eq:total_potential_energy}
\end{align}
Similar to any mechanical system, an equilibrium configuration of the arm is characterized by the (local) minimum of its potential energy landscape, as illustrated in Fig.~\ref{fig:energy_shaping}. We present the mathematical conditions for such an equilibrium next.

\subsubsection{Equilibrium of the arm}		\label{sec:equilibrium_rod}
Consider the time independent activation of a muscle $\muscle$
\begin{equation}\label{eq:static_muscle_activation}
	\muscleactivation(s,t)=\staticmuscleactivation(s),\quad 0\leq s \leq \armlength, ~t\geq0
\end{equation}
Such an activation profile $\staticmuscleactivations=\set{\staticmuscleactivation}$ modifies the stored energy function and potential energy according to \eqref{eq:total_stored_energy} and \eqref{eq:total_potential_energy}, respectively. The change in potential energy alters the equilibrium configuration of the arm. The necessary condition satisfied by the equilibrium is the subject of the following proposition. 

\begin{proposition} \label{prop:static_equilibirum_solution}
	Consider the time independent activation profile \eqref{eq:static_muscle_activation}. Then, the necessary condition for the resulting equilibrium is given by
	\begin{equation}\label{eq:equilibrium_constraint}
		 \frac{\partial W }{\partial\strains}(s, \strains;\staticmuscleactivations)=0, ~~ 0\leq s \leq \armlength
	\end{equation}
\end{proposition}
\begin{proof}
See~\ref{appdx:static_equilibrium_proof}.
\end{proof}

Physically, condition \eqref{eq:equilibrium_constraint} means that total internal loads resulting from passive elasticity and active muscles, are in balance 
	\begin{equation*}
		\internalforces^{\elastic}+{\displaystyle\sum_{\muscle}\internalforces^{\muscle}}=0,\quad
		\internalcouples^{\elastic}+{\displaystyle\sum_{\muscle}\internalcouples^{\muscle}}=0
	\end{equation*}
Henceforth, equation \eqref{eq:equilibrium_constraint} is referred to as the equilibrium constraint. For brevity, we define $P(s, \strains; \staticmuscleactivations) \defined \frac{\partial W }{\partial \strains}(s, \strains; \staticmuscleactivations)$.

\subsection{Design of potential energy: The static problem} \label{sec:energy_shaping_two_steps}
For a fixed time independent activation, the equilibrium is given by \eqref{eq:equilibrium_constraint}. The control design problem then becomes selecting the activation profile that meets the control objective. In applications, additional constrains may arise, for example because of solid obstacles in the environment, which we model as inequalities $\Psi_j(s,\pose(s))\leq0$ for $j=1,2,\dots,N_\text{obs}$ where $N_\text{obs}$ is the total number of obstacles.

We pose the control design as an optimization problem 
		\begin{align}\label{eq:general_static_problem}
		 	\begin{split}
			\min_{\staticmuscleactivations (\cdot), \, \staticmuscleactivations^\muscle (s) \in [0,1]} &\quad \mathsf{J}=\int_0^{\armlength} \Lagrangian(s, \state(s),\staticmuscleactivations(s))\ \dif s + \Phi (\pose(\armlength))\\
			\text{subject to}&\quad \partial_s\pose =\pose\strainmatrix,\quad\state(0)=\state_0,\quad\state(\armlength)~\text{free}\\
			\text{and} &\quad P(s,\strains(s);\staticmuscleactivations(s))=0,~\forall s\in[0, \armlength]\\
			\text{with} & \quad \Psi_j(s, \pose(s)) \leq 0, ~~ j = 1, 2, ..., N_{\text{obs}}
		 	\end{split}
		\end{align}
The form of the objective function $\mathsf{J}$ depends upon each different control objective. Specifically, we write $\mathsf{J}=\mathsf{J}^\text{muscle}(\staticmuscleactivations)+\mathsf{J}^\text{task}(\pose)$, where $\mathsf{J}^\text{muscle}$ models the cost of muscles' activation, and $\mathsf{J}^\text{task}$ depends on the control task. In this paper, we treat the constraints for obstacles $\Psi_j \leq 0$ as soft constraints, following the approach of \cite{chang2021controlling, chang2020energy}. This is done by augmenting the Lagrangian $\Lagrangian$ as $\bar{\Lagrangian}(s, \state,\alpha)=\Lagrangian(s, \state,\staticmuscleactivations)+\sum_j \zeta_jc_j(s,\state)$, where the functions $c_j$ penalize the violation of respective inequalities and $\zeta_j >0$ are some chosen weights. 
 For simplicity of exposition, we ignore the obstacle constraints in the following. 
 
\begin{remark}
Here we inherently assume that the equilibrium constraint $P(s, \strains; \staticmuscleactivations) = 0$ is solvable for all $s$ and for a given $\staticmuscleactivations$. Additionally, we assume the existence of the inverse of the Jacobian matrix $\frac{\partial P}{\partial \strains}$ for all $s$. Thus, according to the implicit function theorem, one may write $\strains = \strains(\staticmuscleactivations)$. This allows us to disregard the equilibrium constraint as an explicit constraint for the optimization problem \eqref{eq:general_static_problem}. These assumptions are also helpful for numerically solving \eqref{eq:general_static_problem}, as explained in \S\,\ref{sec:algorithm}.
\end{remark} 
 
 \begin{remark}
 The optimization problem \eqref{eq:general_static_problem} is sometimes referred to as a bilevel optimization problem or structural optimization problem \cite{colson2007overview, outrata2013nonsmooth}. The equilibrium constraint $P(s, \strains; \staticmuscleactivations) = 0$ is the solution to an optimization problem which we regard as the lower level optimization problem. \ref{appdx:static_equilibrium_proof} and references therein contain details about this lower level problem. This equilibrium constraint is then embedded in the problem of minimizing $\mathsf{J}$, which is regarded as the higher level problem.
 \end{remark}
Solutions to \eqref{eq:general_static_problem} are obtained as follows. First, an enlargement technique is employed so that the classical Pontryagin's Maximum Principle (PMP) \cite{pontryagin1962mathematical, liberzon2011calculus} conditions can be written. 
This is followed by a reduction technique to express the co-state variable in a lower dimensional space for efficient computation. Finally, a numerical algorithm for solving \eqref{eq:general_static_problem} is described in \S\,\ref{sec:algorithm}.

We first notice that the state (the pose of the arm) $\pose$ evolves in $\SE{3}$ according to the equation $\partial_s\pose =\pose\strainmatrix$. However, one can also regard the evolution in $\Mat[4]$, the vector space of all $4\times 4$ real matrices, with the restriction of initial condition $\pose(0) = \pose_0 \in \SE{3}$. This extrinsic viewpoint, known as \textit{enlargement technique}  \cite{brockett1973lie, justh2015enlargement}, is adopted here to obtain the PMP conditions. 
We denote the costate by $\costate\in\Mat[4]$ and we write the control Hamiltonian (or the pre-Hamiltonian) as
	\begin{align}
		\staticHamiltonian(s,\state, \costate, \staticmuscleactivations)=\inner{\costate}{\state\strainmatrix}-\Lagrangian(s, \state, \staticmuscleactivations) 
	\end{align}
where $\inner{\cdot}{\cdot}$ denotes the inner product of two real matrices $\inner{\Matrix{A}}{\Matrix{B}}=\Trace{\Matrix{A}^\transpose\Matrix{B}}$. 
The first order necessary conditions (PMP) for the optimization problem \eqref{eq:general_static_problem} are then given by
	\begin{subequations}\label{eq:PMP}
	\begin{flalign}
	\text{(state)}  \hspace{3.07cm}	
		\partial_s\state&=\frac{\partial\staticHamiltonian}{\partial\costate}=\state\strainmatrix &&  \label{eq:state_evolution}\\
	\text{(costate)} \hspace{3.07cm}	
		\partial_s\costate&=-\frac{\partial\staticHamiltonian}{\partial\state}  && \label{eq:costate_evolution} \\
	\text{(boundary)} \hspace{3cm}	
	 \state(0) &= \state_0, \quad \costate(\armlength) = - \frac{\partial \Phi}{\partial \pose} (\pose(\armlength)) &&
	\label{eq:transversality} \\
	\text{(maximum principle)} \hspace{3cm}	
	 \staticmuscleactivations^{\text{opt}} &= \underset{\staticmuscleactivations (\cdot), \, \staticmuscleactivations^\muscle (s) \in [0,1]}{\arg\!\max} ~~\staticHamiltonian (s, \pose, \costate, \staticmuscleactivations) &&	 \label{eq:maximization_of_Hamiltonian}	
	\end{flalign} 
	\end{subequations}
	where $\staticmuscleactivations^{\text{opt}}$ represents the optimal muscle activations.

We then notice that the costate equation \eqref{eq:costate_evolution} evolves in $\Mat[4]$, a 16 dimensional vector space. However, it is possible to parameterize the costate matrix $\costate$ in terms of two three-dimensional vectors $\internalforces \in \reals^3$ and $\internalcouples \in \reals^3$. This reduction leads to an efficient computation of the optimal control. We present this result as the following proposition. 
%
\begin{proposition} \label{prop:statics_fullaccess}
The costate $\costate$ has the form 
	\begin{subequations}\label{eq:costate_form}
	\begin{equation}
		\costate=\begin{bmatrix}
		\tfrac{1}{2}\orientation\left([\internalcouples]_\times-\symmetricpart\right)&\orientation\internalforces\\
		0&0
		\end{bmatrix}
	\end{equation}
	where the symmetric matrix $\symmetricpart$ is defined as
	\begin{equation}
	\symmetricpart \defined\orientation^\transpose\left(\Lambda+\left(\orientation\internalforces\right)\positions^\transpose+\positions\left(\orientation\internalforces\right)^\transpose \right)\orientation 
	\end{equation}
	with
	\begin{equation}
		\Lambda(s) \defined\int_s^\armlength\frac{\partial \Lagrangian}{\partial\positions}\positions^\transpose+\positions \left(\frac{\partial \Lagrangian}{\partial\positions}\right)^\transpose+\frac{\partial \Lagrangian}{\partial\orientation}\orientation^\transpose+\orientation\left(\frac{\partial \Lagrangian}{\partial\orientation} \right)^\transpose~\dif\bar{s}+\Lambda_{\armlength}
	\end{equation}
	and
	\begin{equation}
		\Lambda_{\armlength}=\left[\frac{\partial \Phi}{\partial\positions}\positions^\transpose+\positions \left(\frac{\partial \Phi}{\partial\positions}\right)^\transpose+\frac{\partial \Phi}{\partial\orientation}\orientation^\transpose+\orientation\left(\frac{\partial \Phi}{\partial\orientation} \right)^\transpose\right]_{s=\armlength}
	\end{equation}
	\end{subequations}
Here, the vectors $\internalforces$ and $\internalcouples$ are the static internal forces and couples in the material frame, respectively. They satisfy the differential equations 
	\begin{subequations}\label{eq:balance_equation}
		\begin{align}
			\partial_s\internalforces&=-\curvatures\times\internalforces + \orientation^\transpose\frac{\partial \Lagrangian}{\partial\positions}\label{eq:balance_equation_n} \\
			\partial_s\internalcouples&=-\curvatures\times\internalcouples-\shears\times\internalforces +\textnormal{vec} \left[\orientation^\transpose\frac{\partial \Lagrangian}{\partial\orientation}-\left(\frac{\partial \Lagrangian}{\partial\orientation} \right)^\transpose\orientation\right]\label{eq:balance_equation_m}
		\end{align}
	and the transversality (boundary) conditions
	\begin{equation} \label{eq:transversality_n_m}
		\internalforces(\armlength)=-\orientation^\transpose(\armlength)\frac{\partial\Phi}{\partial\positions}(\pose(\armlength)),\quad\internalcouples(\armlength)=-\textnormal{vec} \left[\orientation^\transpose(\armlength)\frac{\partial \Phi}{\partial\orientation}(\pose(\armlength))-\left(\frac{\partial \Phi}{\partial\orientation} (\pose(\armlength))\right)^\transpose\orientation(\armlength)\right]
	\end{equation}
	\end{subequations}
\end{proposition}
\begin{proof}
See~\ref{appdx:statics_proof}.
\end{proof}

\noindent
\textbf{Discussion:}
The vectors $\internalforces$ and $\internalcouples$ are identified with the static overall internal forces and couples acting on the rod, respectively. To see this, consider a Lagrangian $\Lagrangian$ that is $\pose$ independent, i.e. the terms $\dfrac{\partial \Lagrangian}{\partial\positions}$ and $\dfrac{\partial \Lagrangian}{\partial\orientation}$ vanish. Then, equations \eqref{eq:balance_equation} are just the static counterparts to the dynamics \eqref{eq:dynamics}, under no external body forces and couples. 
For a $\pose$ dependent Lagrangian $\Lagrangian$, the terms involving $\dfrac{\partial \Lagrangian}{\partial\positions}$ and $\dfrac{\partial \Lagrangian}{\partial\orientation}$ represent the external body forces and couples at the equilibrium. 
Therefore, the parameterization \eqref{eq:costate_form} yields a physical interpretation of the costate variable $\costate$.

\subsubsection{Algorithm for solving the static problem}\label{sec:algorithm} 
\begin{figure*}[!t]
	\centering
	\includegraphics[width=\textwidth , trim = {0 0pt 0pt 0pt}, clip=True]{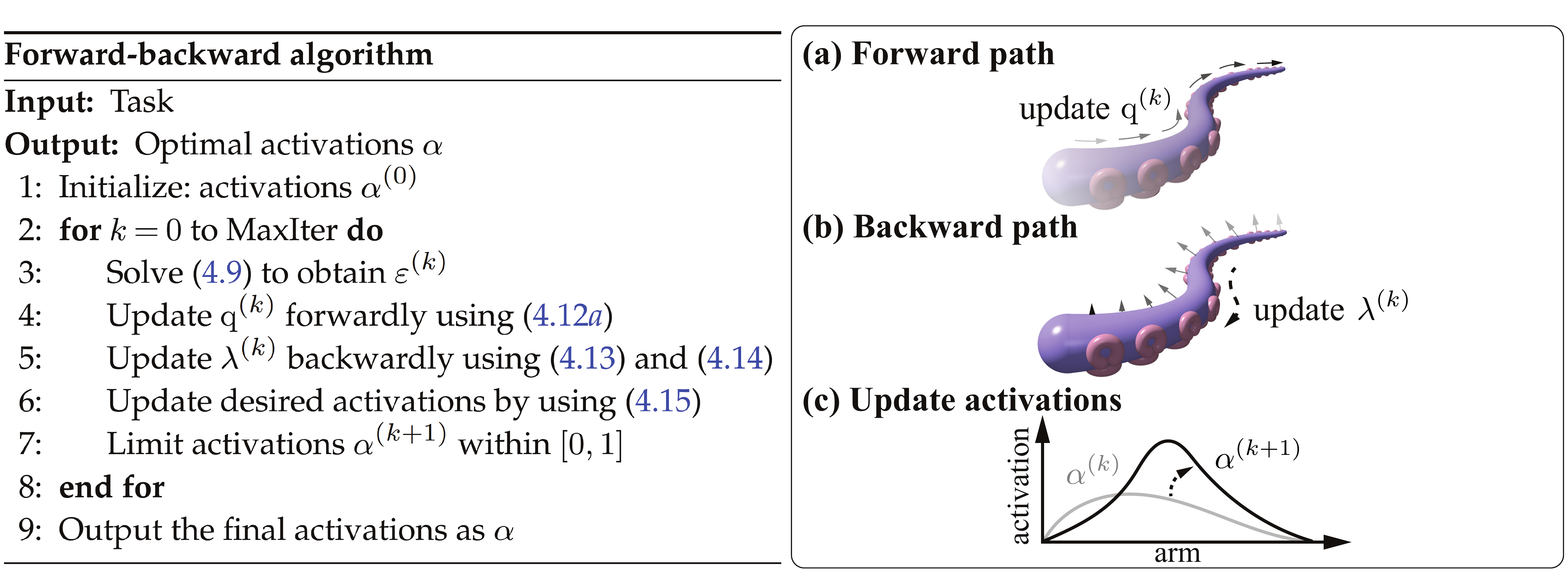}
	\caption{A pseudo code and an illustration of the forward-backward algorithm}
	\label{fig:fb}
\end{figure*}

For a given task, the optimization problem \eqref{eq:general_static_problem} is solved by an iterative forward-backward algorithm to find the desired muscle activation $\staticmuscleactivations$. In each iteration $k$, the equilibrium constraint \eqref{eq:equilibrium_constraint} is first solved pointwise in $s$ to get the strain $\strains^{(k)}$ corresponding to the muscle activation $\staticmuscleactivations^{(k)}$. Next, the state $\pose^{(k)}$ and the co-state $\costate^{(k)}$ are solved by noting the boundary conditions \eqref{eq:transversality} and \eqref{eq:transversality_n_m}. This lets us solve the state equation \eqref{eq:state_evolution} in a \textit{forward} manner from the base to the tip (from $s=0$ to $s= \armlength$), whereas the reduced costate equations \eqref{eq:balance_equation} are solved in a \textit{backward} fashion from the tip to the base (from $s=\armlength$ to $s=0$). The co-state $\costate^{(k)}$ is finally obtained by utilizing the relation \eqref{eq:costate_form}. At the end of each iteration, the control $\staticmuscleactivations^{(k)}$ is updated by using a gradient ascent rule with step size $\eta>0$ in order to maximize the control Hamiltonian $\staticHamiltonian$. The control update rule is written as
\begin{subequations} \label{eq:control_update}
\begin{align}
\staticmuscleactivations^{(k+1)} = \staticmuscleactivations^{(k)} + \eta\left. \frac{\partial \staticHamiltonian}{\partial \staticmuscleactivations} \right|_{\staticmuscleactivations=\staticmuscleactivations^{(k)}}
\end{align}
where the gradient of $\staticHamiltonian$ is computed as
\begin{align}
\frac{\partial \staticHamiltonian}{\partial \staticmuscleactivations} = - \frac{\partial P}{\partial \staticmuscleactivations} \left(\frac{\partial P}{\partial \strains} \right)^{-1} \left( \frac{\partial}{\partial \strains} \inner{\costate}{\pose \strainmatrix}\right) - \frac{\partial \Lagrangian}{\partial \staticmuscleactivations}
\end{align}
\end{subequations}
A detailed discussion of the forward-backward algorithm is found in \cite{chang2020energy, chang2021controlling, mcasey2012convergence}. A brief pseudo code and corresponding illustration are provided in Fig.~\ref{fig:fb}. The value of $\eta$ is listed in Table \ref{tab:parameters}.

\subsection{Energy-shaping control law and dynamic stability} \label{sec:energy_shaping_control_law}
In this section, we argue that the control law given in equation \eqref{eq:static_muscle_activation} is an energy shaping control law for the whole system and discuss its stability properties. We notice the following:
\begin{enumerate}
	\item[(i)] Applying control \eqref{eq:static_muscle_activation} modifies the potential energy from $\potentialenergy^\elastic(\pose)$ to $\potentialenergy(\pose;\staticmuscleactivations)$, which we consider as the desired potential energy $\potentialenergy^\desired$ (see Fig. \ref{fig:energy_shaping}).
	\item[(ii)] The modified Hamiltonian $\Hamiltonian\defined\kineticenergy+\potentialenergy$, under suitable technical conditions, acts as a Lyapunov functional whereby along a solution trajectory
	\begin{equation}\label{eq:Hamiltonian_derivative}
		\frac{\dif}{\dif t}\Hamiltonian(\pose,\momentum)=-\int_0^{\armlength}\abs{\momentums(s,t)}_{\damping\inertia^{-1}}^2~\dif s\leq0
	\end{equation}
\end{enumerate}
Derivation of \eqref{eq:Hamiltonian_derivative} is not straightforward and is shown in \ref{appdx:Hamiltonian_System}. We thus have that the total energy of the system is non-increasing. Finally, an application of the LaSalle's theorem guarantees local asymptotic stability to the largest invariant subset of $\left\lbrace (\pose, \momentums) ~\big| ~ \tfrac{\dif {\Hamiltonian}}{\dif t} = 0 \right\rbrace$, which is indeed the equilibrium point $(\pose^\alpha, 0)$, where $\pose^\alpha$ is the desired static pose. Note that a rigorous functional analytic treatment of the stability requires an analysis of the compact semigroup properties of the closed loop dynamics and an application of the generalized LaSalle's principle for infinite dimensional systems \cite{walker1980dynamical, slemrod1988lasalle}. Such analysis is outside the scope of this paper and will be considered as a future work. 

\begin{remark}
In this paper, we effectively express the muscle actuation term $\Gmuscle \Vector{u}$ as a gradient of a potential energy functional. Thus, the time invariant muscle activation \eqref{eq:static_muscle_activation} restricts the set of allowable desired potential energy functionals. Indeed, under \eqref{eq:static_muscle_activation}, any desired potential energy $\potentialenergy^\desired$ must possess a specific form, namely $\potentialenergy^\desired (\pose)  = \potentialenergy (\pose; \staticmuscleactivations)$, instead of having any functional form. This solves the matching conditions implicitly. Moreover, the static muscle actuation profile $\staticmuscleactivations$ is a solution to the energy shaping control problem. For a finite dimensional setting, this viewpoint is explored in \cite{maschke2000energy}.  
\end{remark}

\section{Arm reaching and grasping in 3D space} \label{sec:numerics}
In this section, two biophysically motivated examples including reaching and grasping are discussed. These well-characterized examples produce 3D motions which are seen as stereotypical motions of an octopus arm~\cite{gutfreund1998patterns, kennedy2020octopus}. Moreover, some of these movements have also been observed in isolated arms~\cite{sumbre2001control}, thereby affording the possibility of experimentally probing them. The numerical simulation environment setup is described in~\S\,\ref{sec:simulation_setup}, and the two example tasks are discussed in~\S\,\ref{sec:exp_reaching} and \S\,\ref{sec:exp_grasping}.


\subsection{Biophysical and numerical setup}\label{sec:simulation_setup}
In order to investigate arm behavior, the explicit Cosserat rod equations \eqref{eq:control_system} are discretized into $N_\text{d}$ connected cylindrical segments and solved numerically by using our open-source, dynamic, and three-dimensional simulation framework \textit{Elastica} \cite{gazzola2018forward, zhang2019modeling, naughton2021elastica}. 

Based on experimental measurements \cite{chang2020energy, chang2021controlling}, the radius profile of the tapered geometry of an octopus arm is modeled as
\begin{align*}
\radius(s)=\radius^{\text{tip}} \frac{s}{\armlength}+\radius^{\text{base}} \frac{\armlength-s}{\armlength}
\end{align*}
where $\radius^{\text{base}}$ and $\radius^{\text{tip}}$ are the radii of the arm at the base and at the tip, respectively. We take the rest configuration of the rod to be straight and of length $\armlength$ (i.e. $\curvatures^\circ\equiv0$, $\shear_1^\circ=\shear_2^\circ\equiv0$ and $\shear_3^\circ\equiv 1$). The muscles are implemented as sources of active internal forces and couples on the arm as described in~\S\,\ref{sec:muscle_force_couple}. The following force length curve $\hillsmodel(\cdot)$ 
is used
\begin{equation*}
	\hillsmodel(\ell)=\max\{3.06 \ell^3 - 13.64 \ell^2 + 18.01 \ell - 6.44,0\}
\end{equation*}
where $\ell > 0$ is the local muscle length. This model is fitted from experimental data~\cite[Fig. 6]{kier2002fast}. 


We also consider the suckers along the octopus arm's side,
which serve a variety of purposes, from sensing chemical stimuli \cite{chase1986chemotactic, wells1963taste} to adhering to surfaces \cite{mather1998octopuses, grasso2008octopus}. This affects the arm's deformation patterns \cite{kennedy2020octopus, wang2022sensory}. For example, during a grasping motion, the arm may need to twist itself so that a large number of suckers are faced towards the object surface. Even though sucker models for sensing or the adhering mechanisms are beyond the scope of this paper, we do consider their one-sided orientation by aligning the director $\director_1 (s)$ with the (virtual) sucker-facing direction, and accounting for this information in designing muscle controls.

Biophysically realistic arm and muscle parameter values are listed in Table~\ref{tab:muscle_model} and also in reference~\cite{chang2020energy}. Parameters for numerical simulations are found in Table~\ref{tab:parameters} with further details in~\cite{gazzola2018forward}.

\begin{table}[tb]
	\centering
	\caption{Parameters for numerical simulation and algorithm}
	\begin{tabular}{clc}
		\rowcolor{black}
		\color{white} Parameter & \color{white} Description & \color{white} Value \\
		\noalign{\smallskip}
		$\armlength$ & rest arm length [cm]& $20$ \\
		$\radius^\text{base}$ & base radius [cm] & $1.2$\\
		$\radius^\text{tip}$ & tip radius [cm] & $0.12$\\
		$E$ & Young's modulus [kPa] & $10$ \\
		$G$ & shear modulus [kPa] & $40/9$ \\
		$\density$ & density [kg/m$^3$] & $1050$ \\
		$\damping_\linearvelocities$ & linear velocity dissipation [1/s] & $0.02$ \\
		$\damping_\angularvelocities$ & angular velocity dissipation [1/s] & $0.02$ \\
		$N_\text{d}$ & discrete number of elements & $100$ \\
		$\Delta t$ & discrete time-step [s] & $10^{-5}$ \\
		$\eta$ & algorithm update step size & $10^{-8}$ \\
		$\mu^\text{pos}$ & position regularization parameter & $10^6$ \\
		$\mu^\text{dir}$ & director regularization parameter & $10^3$ \\
		\hline
	\end{tabular}
	\label{tab:parameters}
\end{table}

\subsection{Reaching a static target}\label{sec:exp_reaching}

The first experiment consists of reaching a stationary target with the tip of the arm in a prescribed, desired orientation.
This mimics the behavior of an octopus fetching food. The form of the objective function in \eqref{eq:general_static_problem} is written as $\mathsf{J}=\mathsf{J}^\text{muscle}+\mathsf{J}^\text{task}$ with 
\begin{equation}
	\mathsf{J}^\text{muscle}(\staticmuscleactivations)=\sum_{\muscle}\int_0^{\armlength}\frac{1}{2}(\staticmuscleactivation(s))^2~\dif s
	\label{eq:J_muscle}
\end{equation}
being the activation cost of muscles. Given the target position $\positions^* \in \R^3$ and the reaching orientation $\orientation^* \in \SO{3}$, we couch the task cost as a function of the distance between desired and achieved tip location and orientation
\begin{equation}
\mathsf{J}^\text{task}=\mathsf{J}^{\text{reach}}(\pose(\armlength))=\frac{\mu^{\text{pos}}}{2}\abs{\positions^*-\positions(\armlength)}^2+\frac{\mu^{\text{dir}}}{2}\norm{\orientation^*-\orientation(\armlength)}_\Frobenius^2
\label{eq:reaching_cost}
\end{equation}
where $\norm{\cdot}_\Frobenius$ denotes the matrix Frobenius norm, and the parameters $\mu^{\text{pos}}, \mu^{\text{dir}} >0$ are regularization weights for position and directors, respectively, and are listed in Table \ref{tab:parameters}.

\begin{figure*}[!t]
	\centering
	\includegraphics[width=\textwidth]{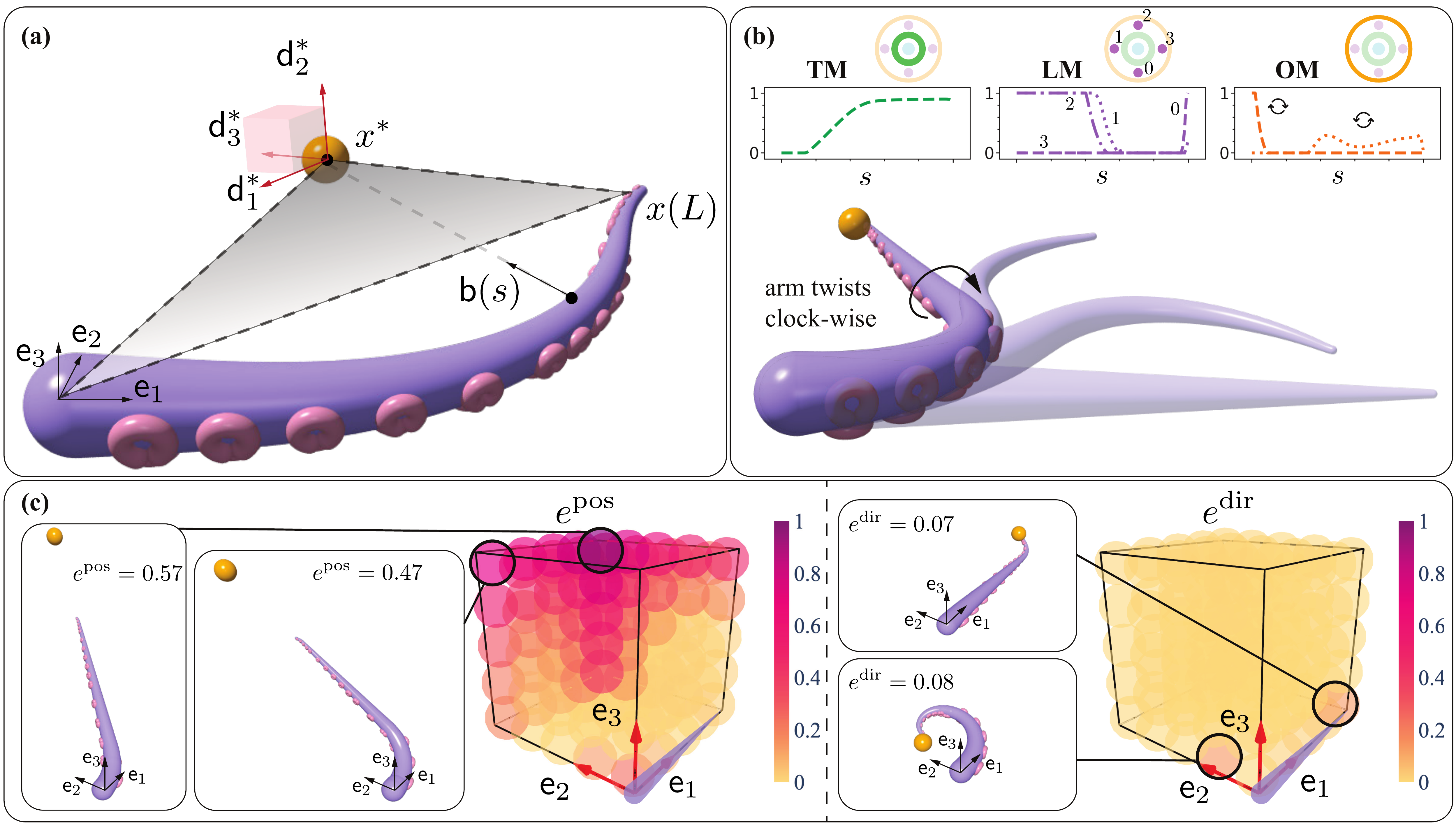}
	\caption{Reaching task. (a) The arm is tasked to reach the target with an orientation that ensures the suckers facing the target,  mimicking real octopus reaching motions.  (b) The target is located at (1,15,6) [cm] and is indicated as an orange sphere.  Muscle activations are shown on top, and the time evolution of the arm is depicted as translucent purple rods.  (c) Different target locations are tested. Two representative cases characterized by large position errors are shown on the left, and two cases characterized by relatively large director errors are shown on the right. The arm on the $\mathsf{e}_1$ axis depicts its initial posture.}
	\label{fig:experiment_reach}
\end{figure*}

The reaching orientation $\orientation^*=\begin{bmatrix}\director_1^*&\director_2^*&\director_3^*\end{bmatrix}$ is determined by
\begin{equation*}
	\director_1^* = \director_2^*\times\director_3^*,\quad
	\director_2^* = \mathsf{b}(0) \times \mathsf{b}(\armlength),\quad
	\director_3^* = \mathsf{b}(\armlength)
\end{equation*}
with
\begin{equation*}
	\mathsf{b}(s)\defined\frac{\positions^*-\positions(s)}{\abs{\positions^*-\positions(s)}}
	\label{eq:definition_of_b}
\end{equation*}
where the vector $\mathsf{b}(s)$ denotes the unit vector pointing from the initial arm configuration to the target position $\positions^*$. The vector $\director_3^*$ aligns with $\mathsf{b}(\armlength)$, $\director_2^*$ is the normal director of the plane spanned by $\mathsf{b}(0)$ and $\mathsf{b}(\armlength)$, and $\director_1^*$ is another unit vector that completes the orthonormal triad (Fig.~\ref{fig:experiment_reach}(a)). This setup is motivated by a common behavior observed in octopuses \cite{kennedy2020octopus}, whereby the arm first points its tip toward the target (i.e. $\director_3(\armlength)$ aligns with $\director_3^*=\mathsf{b}(\armlength)$), and then orients the suckers on the distal end toward the target (i.e. $\director_1(\armlength)$ aligns with $\director_1^*$). 

After formalizing the reaching problem, we proceed with analyzing the ensuing arm behavior. A representative example is illustrated in Fig.~\ref{fig:experiment_reach}(b). The arm, initially at rest in a straight configuration, activates the transverse muscles to extend its distal end, in an attempt of stretching out given the far away target location. 
Two longitudinal muscles near the base are activated so that the arm bends toward to the target. Oblique muscles are also activated, 
so as to twist the arm and reorient the distal suckers toward the target.

Next, we systematically characterize the arm reaching performance by considering
$125$ targets uniformly distributed inside a cube with side length of 20 [cm], as shown in Fig.~\ref{fig:experiment_reach}(c).
The ability of the arm to carry out the task accurately is quantified via
the position and director errors as follows
\begin{equation*}
	e^\text{pos}(\positions(\armlength))=\frac{\abs{\positions^*-\positions(\armlength)}}{\armlength},\quad e^\text{dir}(\orientation(\armlength))=\frac{1}{8}\norm{\orientation^*-\orientation(\armlength)}_{\Frobenius}^2
\end{equation*}
The position error is scaled by the rest length of the arm, and the director error is scaled by 8 which is the maximum value of the squared matrix norm of the difference between two orientation matrices. 
From the position error plot on the left of Fig.~\ref{fig:experiment_reach}(c), we see that the largest errors occur near the corners, opposite to the base of the arm. This is because those targets are physically positioned outside of the reachable workspace of the arm. For these cases, the arm stretches as much as it can in the target direction, but cannot reach it,
as shown in the two insets. For targets very close to the base, it is again difficult for the arm to both bend and twist to properly reach and orient, leading to high errors. We note that director errors are all small, with the maximum error less than 0.1.
This is because the orientation error is evaluated at the very tip of the arm, which is very nimble and can relatively easily reconfigure in space. Such behavior is on display in the right of Fig.~\ref{fig:experiment_reach}(c). The two examples are the ones characterized by the largest director errors, and yet, for all intents and purposes, the tip successfully re-orients as desired. 
Therefore, we conclude that our algorithm is able to compute the muscle activations needed to reliably complete the reaching task.



\subsection{Grasping a static object}\label{sec:exp_grasping}

In the second experiment, the octopus arm is tasked to grasp a target object. This behavior is commonly seen when an octopus tries to reach for a bottle, a stick, or other objects \cite{mather1998octopuses, grasso2008octopus}. In finding the desired static configuration, the object is treated as both an obstacle and a target, so that the arm cannot penetrate its surface but seeks to wrap around it. 

The objective function $\mathsf{J}$ presents a 
decomposition similar to the above described reaching example. 
Here we use the same muscle activation cost $\mathsf{J}^{\text{muscle}}$ \eqref{eq:J_muscle}, while obstacle constraints and task cost are designed as follows.

The target is represented by a closed and convex set $\Object\subset\reals^3$ with its boundary denoted by $\partial\Object$. The non-penetration constraint is modeled by
\begin{equation*}
\Psi(s, \pose(s);\Object)=\radius(s)-\text{dist}(\positions(s),\Object)\leq0
\end{equation*}
where the signed distance function $\text{dist}(\Vector{z},\Object)$ denotes the minimum distance from point $\Vector{z}\in\reals^3$ to the object $\Object$.  Explicitly, the distance function is defined as
	\begin{equation*}
		\text{dist}(\Vector{z},\Object)\defined \left\{\begin{array}{l}
		~~~\min\big\{|\Vector{z}-\Vector{x}'|:\Vector{x}'\in\partial\Object\big\},~\text{if }\Vector{z}\not\in\Object\\
		-\min\big\{|\Vector{z}-\Vector{x}'|:\Vector{x}'\in\partial\Object\big\},~\text{otherwise }
		\end{array}\right.
	\end{equation*}
The negative value of the function $\Psi$ indicates that the arm is not penetrating the boundary.
Other than avoiding penetration, the arm is also expected to wrap around the target. The task cost function is then set to 
\begin{align}\label{eq:grasping_cost}
	\mathsf{J}^\text{task}=\mathsf{J}^\text{grasp}(\pose)=\int_0^\armlength\1(s;s')\left(\mu^\text{pos}\mathsf{J}^\text{pos}(\pose(s),s)+\mu^\text{dir}\mathsf{J}^\text{dir}(\pose(s))\right)~\dif s
\end{align}
where the indicator function is
\begin{equation*}
	\1(s;s')=\left\{\begin{array}{ll}0&\text{if }s\in[0,s')\\1&\text{if }s\in[s',\armlength]\end{array}\right.
\end{equation*}
with $s'=0.3\armlength$ indicating the location at which the arm begins to wrap around the target. The first term $\mathsf{J}^\text{pos}$ minimizes the distance between the arm and the object surface, while the second term $\mathsf{J}^\text{dir}$ induces the suckers to face towards the object for grasping.

\begin{figure*}[!t]
	\centering
	\includegraphics[width=\textwidth]{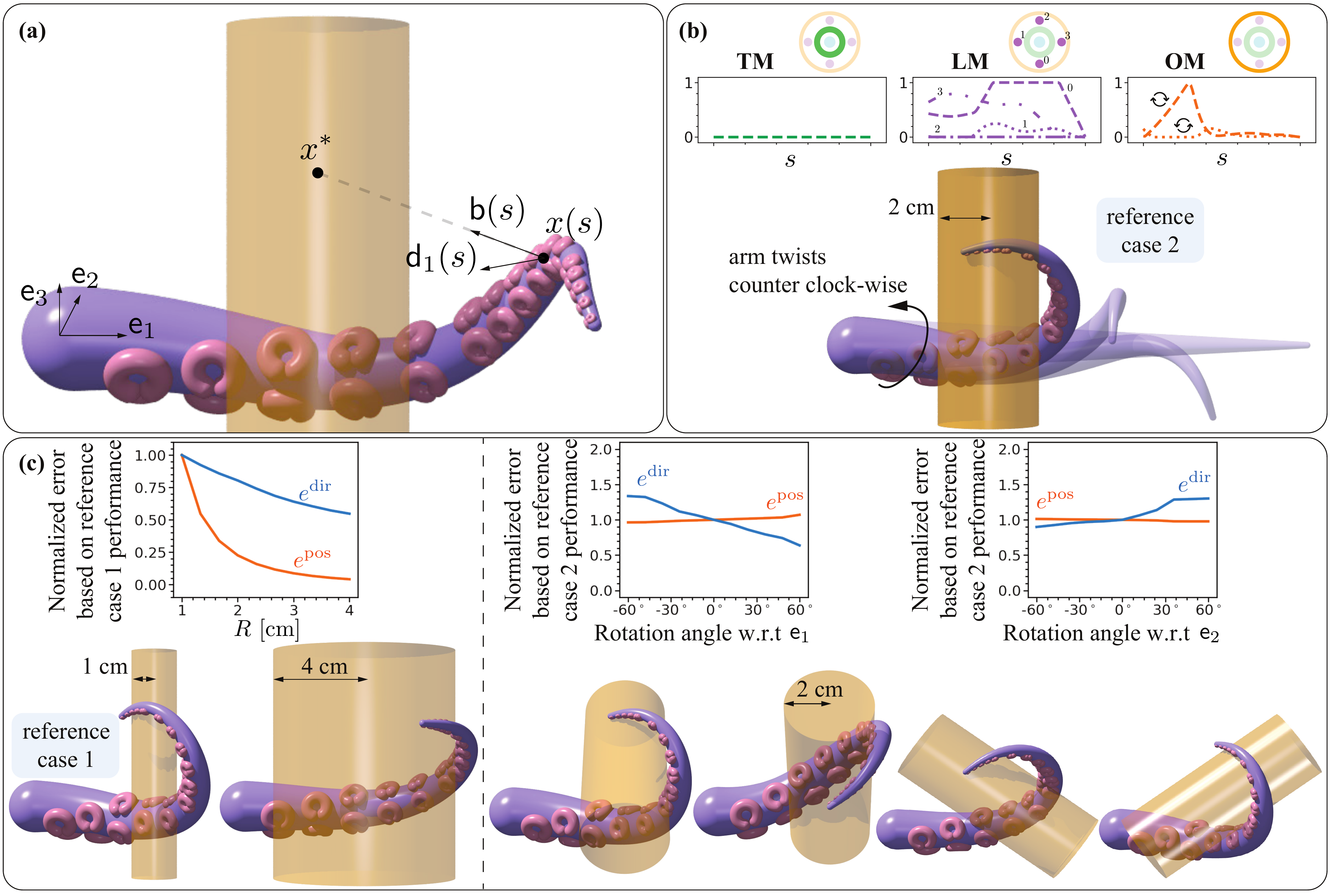}
	\caption{Grasping task. (a) The arm is tasked to wrap around a cylinder of radius $R=\text{2}$ [cm] and height 15 [cm]. The interior point $\positions^*$ is located at (6, -5,2) [cm]. The cylinder is upright, i.e. its orientation aligns with the $\mathsf{e}_3$ axis. The cylinder is shown in translucent orange. (b) Muscle activations are shown on top, and the time evolution of the arm is depicted as translucent purple rods. (c) On the left, the radius of the cylinder $R$ varies from 1 [cm] to 4 [cm]. The normalized (relative to the $R=\text{1}$ [cm] case : reference case 1) error plot and two representative results ($R=\text{1}$ [cm] and $R=\text{4}$ [cm]) are shown. The plots in the middle and on the right consider cylinders rotated about the 
	$\mathsf{e}_1$ and $\mathsf{e}_2$ axes, respectively, keeping $R = 2$ [cm] fixed. The rotation angle varies from $-\text{60}^\circ$ to $\text{60}^\circ$. Normalized (relative to the rotation angle of $\text{0}^\circ$ : reference case 2) error plots and two representative results for $-\text{60}^\circ$ and $\text{60}^\circ$ are presented. }
	\label{fig:experiment_grasp}
\end{figure*}

For this experiment, the arm is initialized straight and is tasked to wrap around an upright solid cylinder target as shown in Fig.~\ref{fig:experiment_grasp}(a).
A point $\positions^* = (6,-5,2)$ [cm] is chosen as an interior point of the cylinder and the cost $\mathsf{J}^\text{pos}$ is designed to minimize surface-to-surface distance
\begin{equation*}
	\mathsf{J}^\text{pos}(\pose(s),s;\positions^*)\defined\frac{1}{2}\left(\abs{\positions^*-\position(s)}-\radius(s)\right)^2
\end{equation*}
while $\mathsf{J}^\text{dir}$ causes $\director_1(s)$ to point towards the cylinder's surface
\begin{equation*}
	\mathsf{J}^\text{dir}(\pose(s);\positions^*)\defined\frac{1}{2}\left(1-\mathsf{b}(s)\cdot\director_1(s)\right)
\end{equation*}

After formalizing the grasping problem, we proceed by demonstrating the arm behavior in Fig.~\ref{fig:experiment_grasp}(b). The arm starts with a straight configuration. The longitudinal and oblique muscles cooperate with each other causing the arm to reconfigure in a spiral shape and wrap around the cylinder. Meanwhile, as 
apparent in Fig.~\ref{fig:experiment_grasp}(b), the arm further modulates twist to approach the cylinder with the suckers side, 
mimicking the grasping behavior. 



For systematic characterization of the grasping performance, we next test different radii and orientation angles of the target cylinder. 
These results are demonstrated in Fig.~\ref{fig:experiment_grasp}(c). To numerically assess the results, we define the following position and director errors
\begin{align*}
	e^\text{pos}(\positions)&=\int_0^\armlength\1(s;s')\frac{1}{2}\left(\frac{\abs{\positions^*-\positions(s)}-\radius(s)}{R}\right)^2\dif s \\
	e^\text{dir}(\orientation)&=\int_0^\armlength\1(s;s')\frac{1}{2}\left(1-\mathsf{b}(s)\cdot\director_1(s)\right)\dif s
\end{align*}
where $R$ is the radius of the cylinder. First, we keep the cylinder upright (orienting toward the $\mathsf{e}_3$ axis) and vary its radius $R$ from $1$ [cm] to $4$ [cm]. As shown in the error plot (on the left of Fig.~\ref{fig:experiment_grasp}(c)), the thinner the cylinder, the higher the position and director errors. In other words, grasping a slender cylinder is harder than 
wrapping around a thick one. This is because larger curvatures of the cylinder surface require greater muscle energy expenditures.
We then keep $R$ fixed at $2$ [cm], and vary its orientation by rotating about the $\mathsf{e}_1$ and $\mathsf{e}_2$ axes, sequentially. This is to test the ability of our control to cope with misalignment. We observe small variability (relative standard deviation of 3.2\% and 1.3\%) in $e^{\text{pos}}$ and larger variability (relative standard deviation of 23.4\% and 14.5\%)  in $e^{\text{dir}}$ for both kinds of rotation. The larger variability in $e^{\text{dir}}$ occurs because the arm prioritizes position error minimization (large $\mu^{\text{pos}}$, see Table~\ref{tab:parameters}).
We therefore conclude that the arm is able to reliably grasp the differently tilted cylinders, however, properly orienting the suckers is harder for certain orientations of the cylinder. 

\section{Conclusion and future work} \label{sec:conclusion}
In this paper, a mathematical model is developed for capturing the three-dimensional motion of a muscular octopus arm. First principle models for all major muscle groups are described in terms of the forces and couples that they generate on a single Cosserat rod representing the arm. Furthermore, the muscles are described via a generalized notion of muscle stored energy functions. The total potential energy of the arm is then expressed by the summation of the passive elastic energy and active muscle potential energy, modulated by muscle control inputs. This is helpful from a control viewpoint since it paves the way for solving an energy shaping control problem. The energy shaping control problem is formulated as a task-specific geometric optimization problem.
First order necessary conditions for such a problem are obtained, and the relationship between the costate variable and the forces and couples acting on the arm is elucidated. Numerical simulations mimicking octopus behavioral experiments demonstrate the engagement of 3D specific deformation modes
(e.g., twist), leading to rich maneuvers. Systematic analyses of the reaching and grasping cases illustrate the ability and robustness of our proposed method. Future research directions include the integration of neuronal activity which drives muscle contractions. Ongoing work also undertakes the sensorimotor control aspect of such a muscular arm.



\aucontribute{
H.-S.C., U.H., and P.G.M. contributed to the theoretical conceptualization of the problem. H.-S.C. and U.H. carried out the modeling of the muscles and the analysis of the energy shaping control methodology. H.-S.C., U.H., M.G. and P.G.M. designed the experiments. H.-S.C. implemented the software. C.-H.S. and N.N. helped interface with the simulation package \textit{Elastica}. N.N. performed the histological analysis of an \textit{Octopus rubescens} arm. C.-H.S. rendered the figures. U.H. and H.-S.C. wrote the manuscript. P.G.M., M.G., and N.N. helped edit the manuscript.}

\competing{The authors declare no competing interests.}

\funding{We gratefully acknowledge financial support from ONR MURI N00014-19-1-2373, NSF EFRI C3 SoRo \#1830881, NSF OAC \#2209322, and ONR N00014-22-1-2569. We also acknowledge computing resources provided by the Extreme Science and Engineering Discovery Environment (XSEDE), which is supported by National Science Foundation grant number ACI-1548562, through allocation TGMCB190004.}



\appendix

\renewcommand\thesection{Appendix \Alph{section}} 
\renewcommand{\thesubsection}{\Alph{section}.\arabic{subsection}}
\renewcommand{\theequation}{\Alph{section}.\arabic{equation}}

\section{Parameter specification in the modeling}\label{appdx:parameters}

\noindent (i) The inertia matrix of a single soft arm takes the form
\begin{equation*}
	\inertia\defined\begin{bmatrix}
	\density\area\identity{3}&0\\0&\density\areamoment
	\end{bmatrix},\quad
	\areamoment=\text{diag}(\areamoment_{11},\areamoment_{22},\areamoment_{33})\defined\frac{\area^2}{4\pi}\begin{bmatrix}
		1&0&0\\
		0&1&0\\
		0&0&2
	\end{bmatrix}
\end{equation*}
where $\identity{3}$ is the $3 \times 3$ identity matrix and $\area=\area(s)$ is the cross sectional area of the rod.\\

\noindent (ii) The shearing and bending rigidity matrices used in the definition of elastic stored potential energy \eqref{eq:elastic_potential_energy} are defined as
\begin{equation*}
	\shearingrigidity\defined\begin{bmatrix}
		G\area&0&0\\
		0&G\area&0\\
		0&0&E\area	
	\end{bmatrix},\quad\bendingrigidity\defined\begin{bmatrix}
		E\areamoment_{11}&0&0\\
		0&E\areamoment_{22}&0\\
		0&0&G\areamoment_{33}
	\end{bmatrix}
\end{equation*}
where $E$ and $G$ are the Young's modulus and shear modulus, respectively.\\

\noindent (iii) Consider the dynamics \eqref{eq:dynamics}. The damping matrix $\damping$ takes the form
\begin{equation*}
	\damping\defined\begin{bmatrix}
	\damping_{\linearvelocities}\identity{3}&0\\0&\damping_{\angularvelocities}\identity{3}
	\end{bmatrix}
\end{equation*}
where $\damping_\linearvelocity, \damping_\angularvelocity >0$  are damping coefficients.  \\

\noindent (iv) Consider the dynamics \eqref{eq:control_system}. The frame transformation velocity matrix $\Upomega$ takes the form
\begin{equation*}
	\Upomega=\begin{bmatrix}
	[\angularvelocities]_\times&0\\ [\linearvelocities]_\times &[\angularvelocities]_\times
	\end{bmatrix}
\end{equation*}

\noindent (v) The muscle actuation model in equation \eqref{eq:control_system} $\Gmuscle \muscleactivations = \sum\limits_{\muscle} \mathsf{G}^\muscle\muscleactivation$ is explicitly written in the following form
\begin{equation*}
	\mathsf{G}^\muscle\muscleactivation=\begin{bmatrix} \partial_s \muscleforces+\curvatures\times\muscleforces\hfill \\
					\partial_s \musclecouples + \curvatures\times\musclecouples +  \shears \times \muscleforces \end{bmatrix}
\end{equation*}
where the models of muscle forces and couples for each type are shown in equation \eqref{eq:transverse_muscle_forces_and_couples},  \eqref{eq:longitudinal_muscle_forces_and_couples}, and  \eqref{eq:oblique_muscle_forces_and_couples}.\\


\newpage
\section{Physiology and modeling of the muscles} \label{appdx:muscle_design}

\subsection{Physiology of an octopus arm}
An \textit{Octopus rubescens} was anesthetized in seawater with 333 nM MgCl concentration. An arm was amputated, fixed in 4\% PFA for 48 hours and transferred to a 1x PBS solution. Hematoxylin and eosin staining was performed using standard protocols. Microscopy was then performed using Hamamatsu’s NanoZoomer slide scanner to achieve full field of view images of the arm’s musculature. Histological cross-sections are provided in Fig.~\ref{fig:muscle_model}.

\subsection{Muscle modeling}
In this section, we describe how each individual muscle is modeled, i.e. its relative position with respect to the center line, strain, length, force, and couple. Note that a tapered arm radius profile $\radius(s)$ is considered as described in~\S\,\ref{sec:simulation_setup}. A summary of this section is given in Table \ref{tab:muscle_model}. \\
	
\noindent\textbf{(i) Transverse Muscle (TM):}\\
\indent These muscles surround the central nerve cord and are arranged perpendicularly to the arm. 
Their relative position to the center line is considered as zero, i.e. $\musclerelativepositions[\TM]=0$ and thus $\muscleshears[\TM] = \shears$ from the definition \eqref{eq:muscleshears}. Since transverse muscles surround the nerve cord, the local length $\musclelength[\TM]$ is proportional to the circumference of the arm. Owing to the constancy of volume, the circumference of the arm is proportional to inverse square root of $\abs{\shears^\TM}$. For simplicity, we then model the local muscle length $\musclelength[\TM]$ as
\begin{equation*}
	\musclelength[\TM]=\sqrt{\frac{\abs{\muscleshears[\TM]_\circ}}{\abs{\muscleshears[\TM]}}}
\end{equation*}

When transverse muscles are activated, the local length of muscles is shortened. Due to conservation of the volume, the arm expands  in the axial direction $\muscletangent[{\TM}]=\muscleshears[\TM]/\abs{\muscleshears[\TM]}$. Thus, the transverse muscles create an extensional force on the tangent direction of the arm.  The internal force is then modeled as (written in the material frame)
\begin{align}\label{eq:transverse_muscle_forces_and_couples}
	\muscleforces[\TM]=-\muscleactivation[\TM]\contractileforce^{\TM}\muscletangent[{\TM}]
\end{align}
and no internal couples are created, i.e. $\musclecouples[\TM]=0$. \\
	
\noindent\textbf{(ii) Longitudinal Muscle (LM):}\\	
\indent These muscles run parallel to the axial nerve cord. Here, we use four to represent these groups of muscle. They are located at the following positions relative to the center line:
\begin{equation*}
	\musclerelativepositions[{\LM_k}]=\beta^{\LM}\begin{bmatrix}\cos(k\pi/2)\\\sin(k\pi/2)\\0\end{bmatrix},\quad k=0,1,2,3
\end{equation*}
where $\beta^{\LM}$ is the ratio of the longitudinal muscle off-center distance with respect to the radius of the arm. Thus, we have the muscle strains
\begin{equation*}
	\muscleshears[{\LM_k}]=\shears+\curvatures\times\musclerelativepositions[{\LM_k}]+\partial_s\musclerelativepositions[{\LM_k}]
\end{equation*}
and muscle length as $\musclelength[{\LM_k}]=\abs{\muscleshears[{\LM_k}]}/\abs{\muscleshears[{\LM_k}]_\circ}$. When each of them is activated, it creates contraction locally on the tangent direction of the muscle and since it is located a distance away from the center line, its contraction force creates couple and thus locally bends the arm. That is
		\begin{equation}\label{eq:longitudinal_muscle_forces_and_couples}
			\muscleforces[{\LM_k}]=\muscleactivation[{\LM_k}]\contractileforce^{\LM_k}\muscletangent[{\LM_k}],\quad\musclecouples[{\LM_k}]=\musclerelativepositions[{\LM_k}]\times\muscleforces[{\LM_k}]
		\end{equation}
with muscle tangent $\muscletangent[{\LM_k}]=\muscleshears[{\LM_k}]/\abs{\muscleshears[{\LM_k}]}$.\\

\noindent\textbf{(iii) Oblique Muscle (OM):}\\
\indent These muscles are similar to the longitudinal muscles, but they do not run parallel to the center line. 
For simplicity, each of these fibers are considered to position like a helical curve with $N_\cycle$ number of cycles around the center line (see Fig.~\ref{fig:muscle_model}). Two types of winding are considered here: rotating clock-wise and counter clock-wise. For each type of winding, they are modeled by four muscles and we write their relative positions as
\begin{align*}
	\musclerelativepositions[{\OM^+_k}]=\beta^{\OM}\begin{bmatrix}\cos(k\pi/2+2\pi N_\cycle s/\armlength)\\\sin(k\pi/2+2\pi N_\cycle s/\armlength)\\0\end{bmatrix},\quad k=0,1,2,3, ~ N_\cycle = 6
\end{align*}
for clock-wise oblique muscles and 
\begin{align*}
	\musclerelativepositions[{\OM^-_k}]=\beta^{\OM}\begin{bmatrix}\cos(k\pi/2-2\pi N_\cycle s/\armlength)\\\sin(k\pi/2-2\pi N_\cycle s/\armlength)\\0\end{bmatrix},\quad k=0,1,2,3, ~ N_\cycle = 6
\end{align*}
for counter clock-wise oblique muscles. Note that $\beta^{\OM}$ is the ratio of the longitudinal muscle off-center distance with respect to the radius of the arm. The muscle strains are then calculated by ($\OM^+_k$ for example)
\begin{align*}
	\muscleshears[{\OM^+_k}]=\shears+\curvatures\times\musclerelativepositions[{\OM^+_k}]+\partial_s\musclerelativepositions[{\OM^+_k}]
\end{align*}
and the muscle length $\musclelength[{\OM^+_k}]=\abs{\muscleshears[{\OM^+_k}]}/\abs{\muscleshears[{\OM^+_k}]_\circ}$. Muscle force and couple are then modeled using the similar equation as the longitudinal muscle. Within each group of differently wound oblique muscle groups, the four fibers are activated together locally, i.e. only two controls: $\muscleactivation[{\OM^+}](s,t)$ and $\muscleactivation[{\OM^-}](s,t)$ control two groups of oblique muscles. For example, for the $\OM^+$ group, 
\begin{equation}\label{eq:oblique_muscle_forces_and_couples}
\begin{aligned}
			\muscleforces[{\OM^+_k}]=\muscleactivation[{\OM^+}] \contractileforce^{{\OM^+_k}}\muscletangent[{\OM^+_k}],\quad
			\musclecouples[{\OM^+_k}]= \musclerelativepositions[{\OM^+_k}]\times \muscleforces[{\OM^+_k}]
		\end{aligned}		
		\end{equation}
with muscle tangent $\muscletangent[{\OM^+_k}]=\muscleshears[{\OM^+_k}]/\abs{\muscleshears[{\OM^+_k}]}$. Similar equations follow for the counter clock-wise rotated muscle group. Cumulatively, one muscle group mainly contributes to twist the arm, e.g. the $\OM^+$ group provides counter clock-wise twisting.



\section{Proof of Proposition \ref{prop:muscle_energy}} \label{appdx:muscle_energy_proof}
\begin{proof}
Indeed, define a vector valued function 
\begin{equation*}
	\Vector{g}^\muscle (s, \Vector{y}) \defined  \left\{
\begin{aligned}
&\quad~\contractileforce^\muscle(s, \musclelength(\Vector{\abs{y}}))\frac{\Vector{y}}{\abs{\Vector{y}}}, &&\text{if } ~\muscle \in \{ \LM, \OM\} \\
&-\contractileforce^\muscle(s, \musclelength(\Vector{\abs{y}}))\frac{\Vector{y}}{\abs{\Vector{y}}}, &&\text{if } ~ \muscle = \TM
\end{aligned}\right.	
\end{equation*}
with $\Vector{y}\in\reals^3, \Vector{y} \neq 0$, and a scalar function $\mu^\muscle (s, \Vector{y})$ as follows
\begin{equation*}
\mu^\muscle (s, \Vector{y}) \defined \left\{
\begin{aligned}
&\int_{\musclelength_\intrinsic}^{\musclelength(\abs{\Vector{y}})} \maxmusclestress_{\text{max}}\area^\muscle (s)\hillsmodel(\ell)\abs{\muscleshears_\intrinsic}~\dif \ell, &&\text{if } ~\muscle \in \{ \LM, \OM\} \\
&\int_{\musclelength_\intrinsic}^{\musclelength(\abs{\Vector{y}})} 2\maxmusclestress_{\text{max}}\area^\muscle (s)\hillsmodel(\ell)\abs{\muscleshears_\intrinsic}~\frac{ \dif \ell}{\ell^3}, &&\text{if } ~ \muscle = \TM
\end{aligned}\right.
\end{equation*}
where ${\musclelength_\intrinsic} = \musclelength(\abs{\Vector{\muscleshears_\intrinsic}})$.
Then through a straightforward calculation it is clear that,  $\Vector{g}^\muscle (s, \Vector{y}) = \nabla_{\Vector{y}} \, \mu^\muscle (s, \Vector{y})$ 
for all muscle $\muscle$, where we have used the definitions of the muscle length function $\musclelength(\cdot)$ as per Table~\ref{tab:muscle_model}. This means the line integral in \eqref{eq:muscle_stored_energy} is path-independent and hence we can write
\begin{align*}
W^\muscle (s, \muscleshears) = \mu^\muscle (s, \muscleshears)
\end{align*}
This implies
\begin{align*}
\frac{\partial W^\muscle}{\partial \muscleshears}(s,\muscleshears) = \frac{\partial \mu^\muscle}{\partial \muscleshears} = \Vector{g}^\muscle (s, \muscleshears) 
\end{align*}
Then it is clear that 
\begin{align*}
 u^\muscle \frac{\partial W^\muscle}{\partial \shears} &= u^\muscle \frac{\partial \shears^\muscle}{\partial \shears}\frac{\partial W^\muscle}{\partial \shears^\muscle} = \pm u^\muscle \contractileforce^\muscle \frac{{\muscleshears}}{\abs{{\muscleshears}}} =  \muscleforces, \\
\text{and} \quad 
u^\muscle \frac{\partial W^\muscle}{\partial \curvatures} &= u^\muscle \frac{\partial \shears^\muscle}{\partial \curvatures}\frac{\partial W^\muscle}{\partial \shears^\muscle} = \musclerelativepositions \times \muscleforces = \musclecouples. 
\end{align*}	
\end{proof}

\section{Proof of Proposition \ref{prop:static_equilibirum_solution}}

\begin{proof}\label{appdx:static_equilibrium_proof}
For a given set of static muscle activation profile $\staticmuscleactivations$, the arm is subjected to a fixed-free type boundary condition. For such a scenario, static equilibria of the arm are obtained by solving an optimization problem \cite{bretl2014quasi, chang2020energy, chang2021controlling,till2017elastic} as follows
\begin{equation}\label{eq:static_problem_given_alpha}
\begin{aligned}
			\min_{\strains(\cdot)}&\quad\potential =\int_0^{\armlength} \storedenergy(s, \strains; \staticmuscleactivations) ~ \dif s \\
			\text{s. t.}&\quad \partial_s\pose =\pose\strainmatrix,\quad\pose(0)=\pose_0,\quad\pose(\armlength)\text{ free}
\end{aligned}
\end{equation}
Here $\storedenergy$ is the total stored energy function as introduced in \eqref{eq:total_stored_energy}. Any minimizer of the problem \eqref{eq:static_problem_given_alpha}, denoted by $\strains$, must satisfy the necessary conditions of optimality.  Let the control Hamiltonian be
\begin{align*}
	H(\state, \strains, \costate;\staticmuscleactivations)=\inner{\costate}{\state\strainmatrix}-\storedenergy(s, \strains;\staticmuscleactivations)
\end{align*}
	where $\costate\in\Mat[4]$ is the costate. The Hamilton's equations are then given by the PMP condition
	\begin{align*}
		\partial_s\state&=\frac{\partial H}{\partial\costate}=\state\strainmatrix,\quad\state(0)=\state_0\\
		\partial_s\costate&=-\frac{\partial H}{\partial\state}=-\costate\strainmatrix^\transpose,\quad\costate(\armlength)=0
	\end{align*} 
	and the optimal strains satisfy the maximization of the Hamiltonian
	\begin{equation}
	   \frac{\partial H}{\partial \strain} = 0
		\label{eq:maximize_hamiltonian}
	\end{equation}
Indeed, from the transversality condition $\costate (\armlength) = 0$ (due to the free boundary condition at the tip) and the evolution equation $\partial_s \costate = - \costate \strainmatrix^\transpose$, we see that $\costate (s) \equiv 0$ for all $s$. Then, the definition of the total stored energy function \eqref{eq:total_stored_energy} and Proposition~\ref{prop:muscle_energy} readily yield equation \eqref{eq:equilibrium_constraint} from maximization of the Hamiltonian \eqref{eq:maximize_hamiltonian}.
\end{proof}

\section{Proof of Proposition \ref{prop:statics_fullaccess}} \label{appdx:statics_proof}
\begin{proof}
	At the outset, we first define a transformation of the costate as 
	\begin{equation*}
		\transformedcostate=\state^\transpose\costate
	\end{equation*}
	and then its evolution and the transversality condition are derived as
	\begin{equation}\label{eq:transformedcostate_evolution}
		\partial_s\transformedcostate=\liebracket{\strainmatrix^\transpose}{\transformedcostate}+\state^\transpose\frac{\partial \Lagrangian}{\partial\state},\quad\transformedcostate(\armlength)=-\pose^\transpose (\armlength) \frac{\partial \Phi}{\partial \pose}(\pose(\armlength))
	\end{equation}
	where $\liebracket{\Matrix{A}}{\Matrix{B}}=\Matrix{A}\Matrix{B}-\Matrix{B}\Matrix{A}$ is the standard matrix commutator. It is then clear that solving the costate equation \eqref{eq:costate_evolution} is equivalent to solving \eqref{eq:transformedcostate_evolution}, since once the $\transformedcostate$ is solved, the costate $\costate$ is then given by
	\begin{equation}\label{eq:transformed_costated_to_costate}
		\costate=(\state^\transpose)^\inverse\transformedcostate
	\end{equation}
	where the inverese always exists. 
	Next, we consider the following form of the transformed costate 
	\begin{equation*}
		\transformedcostate=\begin{bmatrix}
		\transformedcostateA&\transformedcostateC\\[5pt]
		\transformedcostateB^\transpose&\transformedcostateD
		\end{bmatrix}
	\end{equation*}
	where $\transformedcostateA(s)\in\reals^{3\times 3}$, $\transformedcostateB(s)\in\reals^3$, $\transformedcostateC(s)\in\reals^3$ and $\transformedcostateD(s)\in\reals$. Based on this partition, \eqref{eq:transformedcostate_evolution} becomes
	\begin{subequations}
		\begin{align}
			\partial_s\transformedcostateA&=\liebracket{[\curvatures]_\times^\transpose}{\transformedcostateA}-\transformedcostateC\shears^\transpose+\orientation^\transpose\frac{\partial \Lagrangian}{\partial\orientation},&&\transformedcostateA(\armlength)=-\orientation^\transpose(\armlength)\frac{\partial \Phi}{\partial\orientation}(\pose(\armlength))\label{eq:transformedcostateA_evolution}\\
			\partial_s\transformedcostateB^\transpose&=\shear^\transpose\transformedcostateA-\transformedcostateB^\transpose[\curvatures]_\times^\transpose-\transformedcostateD\shears^\transpose+\positions^\transpose\frac{\partial \Lagrangian}{\partial\orientation},&&\transformedcostateB^\transpose(\armlength)=-\positions^\transpose(\armlength)\frac{\partial \Phi}{\partial\orientation}(\pose(\armlength))\label{eq:transformedcostateB_evolution}\\
			\partial_s\transformedcostateC&=[\curvatures]_\times^\transpose\transformedcostateC+\orientation^\transpose\frac{\partial \Lagrangian}{\partial\positions},&&\transformedcostateC(\armlength)=-\orientation^\transpose(\armlength)\frac{\partial \Phi}{\partial\positions}(\pose(\armlength))\label{eq:transformedcostateC_evolution}\\
			\partial_s\transformedcostateD&=\shears^\transpose\transformedcostateC+\positions^\transpose\frac{\partial \Lagrangian}{\partial\positions},&&\transformedcostateD(\armlength)=-\positions^\transpose(\armlength)\frac{\partial \Phi}{\partial\positions}(\pose(\armlength))\label{eq:transformedcostateD_evolution}
		\end{align}
	\end{subequations}
	Now, we define $\internalforces :=\transformedcostateC$ and rewrite \eqref{eq:transformedcostateC_evolution} as
	\begin{equation*}
		\partial_s\internalforces=-\curvatures\times\internalforces+\orientation^\transpose\frac{\partial \Lagrangian}{\partial\positions}
	\end{equation*}
	with the transversality condition
	\begin{equation*}
	\internalforces(\armlength)=-\orientation^\transpose(\armlength)\frac{\partial \Phi}{\partial\positions}(\pose(\armlength))
	\end{equation*}
	By integrating both sides of \eqref{eq:transformedcostateD_evolution} and utilizing \eqref{eq:transformedcostateC_evolution}, we have the following form of $\transformedcostateD$ 
	\begin{equation*}
		\transformedcostateD=\positions^\transpose\orientation\internalforces
	\end{equation*}
	To find the form of $\transformedcostateA$, we first can write it as a combination of the symmetric part and skew-symmetric part and derive their corresponding evolution. The skew-symmetric part's evolution is given by
	\begin{equation*}
			\partial_s(\transformedcostateA-\transformedcostateA^\transpose)=\liebracket{\transformedcostateA-\transformedcostateA^\transpose}{[\curvatures]_\times}-\left(\internalforces\shears^\transpose-\shears\internalforces^\transpose\right)\\
			+\left(\orientation^\transpose\frac{\partial \Lagrangian}{\partial\orientation}-\left(\frac{\partial \Lagrangian}{\partial\orientation}\right)^\transpose\orientation\right)
	\end{equation*}
	Defining $[\internalcouples]_\times := \transformedcostateA-\transformedcostateA^\transpose$, for some $m(s) \in \R^3$, we have
	\begin{equation*}
		\partial_s\internalcouples=-\curvatures\times\internalcouples-\shears\times\internalforces+\text{vec}\left[\orientation^\transpose\frac{\partial \Lagrangian}{\partial\orientation}-\left(\frac{\partial \Lagrangian}{\partial\orientation}\right)^\transpose\orientation\right]
	\end{equation*}
	with the transversality condition
	\begin{equation*}
		\internalcouples(\armlength)=-\text{vec}\left[\orientation^\transpose\frac{\partial \Phi}{\partial\orientation}-\left(\frac{\partial \Phi}{\partial\orientation}\right)^\transpose\orientation\right]
	\end{equation*}
	The evolution of the symmetric part of $\transformedcostateA$ is given by
	\begin{equation*}
		\partial_s(\transformedcostateA+\transformedcostateA^\transpose)=\liebracket{\transformedcostateA+\transformedcostateA^\transpose}{[\curvatures]_\times}-\left(\internalforces\shears^\transpose+\shears\internalforces^\transpose\right)+\left(\orientation^\transpose\frac{\partial \Lagrangian}{\partial\orientation}+\left(\frac{\partial \Lagrangian}{\partial\orientation}\right)^\transpose\orientation\right)
	\end{equation*}
	By integrating both sides of the equation and after some direct calculations, we have
	\begin{equation*}
		\symmetricpart=\orientation^\transpose\left(\Lambda+(\orientation\internalforces)\positions^\transpose+\positions\left(\orientation\internalforces\right)^\transpose \right)\orientation 
	\end{equation*}
	where we denote $\symmetricpart :=-\left(\transformedcostateA+\transformedcostateA^\transpose\right)$ and $\Lambda$ is defined as
	\begin{equation*}
		\Lambda(s) \defined\int_s^\armlength\frac{\partial \Lagrangian}{\partial\positions}\positions^\transpose+\positions \left(\frac{\partial \Lagrangian}{\partial\positions}\right)^\transpose+\frac{\partial \Lagrangian}{\partial\orientation}\orientation^\transpose+\orientation\left(\frac{\partial \Lagrangian}{\partial\orientation} \right)^\transpose~\dif\bar{s}+\Lambda_{\armlength}
	\end{equation*}
	where
	\begin{equation*}
		\Lambda_{\armlength} = \left[\frac{\partial \Phi}{\partial\positions}\positions^\transpose+\positions \left(\frac{\partial \Phi}{\partial\positions}\right)^\transpose+\frac{\partial \Phi}{\partial\orientation}\orientation^\transpose+\orientation\left(\frac{\partial \Phi}{\partial\orientation} \right)^\transpose\right]_{s=\armlength}
	\end{equation*}
	Finally we write
	\begin{equation*}
		\transformedcostateA=\tfrac{1}{2}([\internalcouples]_\times-\symmetricpart)
	\end{equation*}
	Lastly, by integrating both sides of \eqref{eq:transformedcostateB_evolution}, we obtain
	\begin{equation*}
		\transformedcostateB^\transpose=\positions^\transpose\orientation\transformedcostateA
	\end{equation*}
	Now, we already have the form of the transformed costate $\transformedcostate$. To find the form of the costate, we calculate \eqref{eq:transformed_costated_to_costate} and have 
	\begin{equation*}
		\costate=(\state^\transpose)^\inverse\transformedcostate=\begin{bmatrix}\orientation&0\\[5pt]-\positions^\transpose\orientation&1\end{bmatrix}\begin{bmatrix}\transformedcostateA&\internalforces\\[5pt]\positions^\transpose\orientation\transformedcostateA&\positions^\transpose\orientation\internalforces\end{bmatrix}\\
		=\begin{bmatrix}\tfrac{1}{2}\orientation([\internalcouples]_\times-\symmetricpart)&\orientation\internalforces\\[5pt]0&0\end{bmatrix}
	\end{equation*}
\end{proof}

\section{Time derivative of the Hamiltonian -- \\ derivation of \eqref{eq:Hamiltonian_derivative}}\label{appdx:Hamiltonian_System}
	
	Recall that the Hamiltonian of the rod system is $\Hamiltonian(\pose,\momentums)\defined\kineticenergy(\momentums)+\potentialenergy(\pose)$
	and its time derivative is
	\begin{equation}\label{eq:Hamiltonian_time_derivative}
		\begin{aligned}
			\frac{\dif}{\dif t}\Hamiltonian(\pose,\momentums)=&\int_0^\armlength
				(\inertia^{-1}\momentums)^\transpose(\partial_t\momentums)+(\partial\storedenergy/\partial\strains)^\transpose(\partial_t\strains)~\dif s\\
			=&\int_0^\armlength
				\linearvelocities^\transpose\partial_t(\density\area\linearvelocities)
				+\angularvelocities^\transpose\partial_t(\density\areamoment\angularvelocities)
				+\internalforces^\transpose(\partial_t\shears)
				+\internalcouples^\transpose(\partial_t\curvatures)~\dif s
		\end{aligned}
	\end{equation}
	We first focus on the first and the third terms
	\begin{align*}
		\int_0^\armlength
				\linearvelocities^\transpose\partial_t(\density\area\linearvelocities)
				+\internalforces^\transpose(\partial_t\shears)~\dif s
		=&\int_0^\armlength\nonumber
				\linearvelocities^\transpose\partial_t(\density\area\linearvelocities)
				+(\orientation\internalforces)^\transpose\big(\partial_s(\orientation\linearvelocities)-(\partial_t\orientation)\shears\big)~\dif s\nonumber\\
		=&\int_0^\armlength
				\linearvelocities^\transpose\partial_t(\density\area\linearvelocities)
				-\big(\orientation^\transpose\partial_s(\orientation\internalforces)\big)^\transpose\linearvelocities-\internalforces^\transpose\orientation^\transpose\partial_t\orientation\shears~\dif s\nonumber\\
		=&\int_0^\armlength
				-\damping_\linearvelocities\abs{\density\area\linearvelocities}_{(\density\area)^{-1}}^2-\internalforces\cdot(\angularvelocities\times\shears)~\dif s
	\end{align*}
	Next, for the second and the forth terms in \eqref{eq:Hamiltonian_time_derivative}, we have
	\begin{align*}
		\int_0^\armlength
				\angularvelocities^\transpose\partial_t(\density\areamoment\angularvelocities)
				+\internalcouples^\transpose(\partial_t\curvatures)~\dif s
		=&\int_0^\armlength\nonumber
				\angularvelocities^\transpose\partial_t(\density\areamoment\angularvelocities)
				+(\orientation\internalcouples)^\transpose\partial_s(\orientation\angularvelocities)~\dif s\nonumber\\
		=&\int_0^\armlength
				\angularvelocities^\transpose\partial_t(\density\areamoment\angularvelocities)
				-\big(\orientation^\transpose\partial_s(\orientation\internalcouples)\big)^\transpose\angularvelocities~\dif s\nonumber\\
		=&\int_0^\armlength
				\angularvelocities\cdot(\shears\times\internalforces)-\damping_\angularvelocities\abs{\density\areamoment\angularvelocities}_{(\density\areamoment)^{-1}}^2~\dif s
	\end{align*}
Together, the time derivative of the Hamiltonian \eqref{eq:Hamiltonian_time_derivative} follows
	\begin{equation*}
	\frac{\dif}{\dif t}\Hamiltonian(\pose,\momentum)=-\int_0^{\armlength}\abs{\momentums}_{\damping\inertia^{-1}}^2~\dif s\leq0
	\end{equation*}
	with damping coefficients $\damping_\linearvelocities,\damping_\angularvelocities>0$.

\bibliographystyle{RS}
\bibliography{reference}     

\end{document}